\documentclass[12pt, draftclsnofoot,onecolumn]{IEEEtran}
\usepackage{dsfont}
\usepackage{amssymb}
\usepackage{subfig}
\usepackage{graphicx, amssymb, amsmath}
\usepackage{extarrows}
\usepackage{algorithm} 
\usepackage{algorithmic} 
\usepackage{multirow} 
\usepackage{color}
\emergencystretch=\maxdimen
\IEEEoverridecommandlockouts
\begin{document}

\title{Energy-Aware Traffic Offloading for Green Heterogeneous Networks}

\author{Shan~Zhang,~\IEEEmembership{Student~Member,~IEEE,}
        Ning~Zhang,~\IEEEmembership{Member,~IEEE,}
        Sheng~Zhou,~\IEEEmembership{Member,~IEEE,}
        Jie~Gong,~\IEEEmembership{Member,~IEEE,}
        Zhisheng~Niu,~\IEEEmembership{Fellow,~IEEE,}
        and~Xuemin~(Sherman)~Shen,~\IEEEmembership{Fellow,~IEEE}
\thanks{This work is sponsored in part by the National Basic Research Program of China (973 Program: No. 2012CB316001), the National Science Foundation of China (NSFC) under grant No. 61571265, No. 61321061, No. 61401250, and No. 61461136004, Hitachi R\&D Headquater, and NSERC of Canada. Corresponding author: Ning Zhang.}
\thanks{Shan~Zhang, Sheng~Zhou, and Zhisheng~Niu are with Tsinghua National Laboratory for Information Science and Technology, Tsinghua University, Beijing, 100084, P.R. China (Email: zhangshan11@mails.tsinghua.edu.cn, \{sheng.zhou, niuzhs\}@tsinghua.edu.cn).}
\thanks{Ning~Zhang and Xuemin~(Sherman)~Shen are with the Department of Electrical and Computer Engineering, University of Waterloo, 200 University Avenue West, Waterloo, Ontario, Canada, N2L 3G1 (Email: n35zhang@uwaterloo.ca, xshen@bbcr.uwaterloo.ca).}
\thanks{Jie Gong is with the School of Data and Computer Science, Sun Yat-Sen University, Guangzhou, P.R.China, 510006 (Email: xiaocier04@gmail.com).}
\thanks{Part of this work has been presented in IEEE GLOBECOM~2015 \cite{mine_GC_EH}.}
}

\maketitle

\begin{abstract}

With small cell base stations (SBSs) densely deployed in addition to conventional macro base stations (MBSs), the heterogeneous cellular network (HCN) architecture can effectively boost network capacity.
To support the huge power demand of HCNs, renewable energy harvesting technologies can be leveraged.
In this paper, we aim to make efficient use of the harvested energy for on-grid power saving while satisfying the quality of service (QoS) requirement.
To this end, energy-aware traffic offloading schemes are proposed, whereby user associations, ON-OFF states of SBSs, and power control are jointly optimized according to the statistical information of energy arrival and traffic load.
Specifically, for the single SBS case, the power saving gain achieved by activating the SBS is derived in closed form, based on which the SBS activation condition and optimal traffic offloading amount are obtained.
Furthermore, a two-stage energy-aware traffic offloading (TEATO) scheme is proposed for the multiple-SBS case, considering various operating characteristics of SBSs with different power sources.
Simulation results demonstrate that the proposed scheme can achieve more than 50\% power saving gain for typical daily traffic and solar energy profiles, compared with the conventional traffic offloading schemes.
\end{abstract}


\section{Introduction}


Mobile data traffic is predicted to have a 1000-fold growth by 2020, compared with that in 2010 due to the proliferation of wireless devices and emerging multimedia services \cite{Qualcomm}. To accommodate such a huge amount of mobile traffic, small cell base stations (SBSs) are expected to be densely deployed
to offload traffic from the conventional macro base stations (MBSs), forming heterogeneous cellular networks (HCNs) \cite{NZhang_cloud_5G}.
Despite the high network capacity, the dense SBSs also require huge power supply, causing heavy burdens to both the network operators and the power grid \cite{Ismail14_green_survey_Zhuang}.

To deal with the cumbersome energy consumption, energy harvesting (EH) technology can be introduced into HCNs.
Specifically, the emerging EH-SBSs, which are equipped with EH devices (like solar panels or wind turbines) and exploit renewable energy as supplementary or alternative power sources, have received great attentions from both academia and industry \cite{JSAC_overview_5G}.
The possibility and reliability of self-powered cellular networks are investigated in \cite{Jeffery_fundamental}.
The system costs of EH-BSs are evaluated in \cite{green_cellular_cost}, suggesting that renewable energy can be a sustainable and economical alternative if properly managed.
Besides, EH as well as the mmWave-based high-speed wireless backhaul enable SBSs deployed in a distributed plug-and-play manner without wired connections, making network deployment more flexible and cost-effective \cite{Jeffery_fundamental}.
Telecommunication equipment vendors have designed and built green energy powered base stations (BSs) in rural areas, and over 400,000 off-grid BSs will be deployed by 2020 \cite{Navigant_report}.

However, EH poses significant challenges for network operation and management.
Firstly, unlike the conventional on-grid power supply, renewable energy arrives randomly depending on the weather condition.
Secondly, the traffic load is non-uniformly distributed in both spatial and temporal domains, which may not be in accordance with the harvested energy status \cite{Tao_Magazine_RE} \cite{Zhongming}.
Thus, energy waste and service outage could happen without effective energy management strategies, degrading the system reliability and sustainability.
Thirdly, diverse types of SBSs with different energy sources will coexist, including on-grid conventional SBSs (CSBSs), off-grid SBS powered solely by renewable energy (RSBSs), and hybrid SBSs (HSBSs) jointly powered by harvested energy and power grid.
Their different operating characteristics should be also taken into consideration for the design of network management schemes.
Therefore, how to fully utilize the harvested energy to minimize on-grid power saving while satisfying the quality of service (QoS) requirement is a critical issue.

In the literature, a flurry of research work has been reported to improve the utilization of harvested energy \cite{EH_single_link_2}-\cite{Sheng_EH_sleep} \cite{Tao_ICE_long}-\cite{mine_GC_EH}.
The optimal link-level transmission strategies are studied in \cite{EH_single_link_2}, by applying queueing theory to model the random arrival of data and energy.
For BS-level operations, online and offline resource allocation schemes are proposed to maximize the energy efficiency for the OFDMA BS system jointly powered by harvested energy and power grid \cite{Derrick_single_BS_OFDMA}, and dynamic cell deactivation is further considered in \cite{Sheng_EH_sleep}.
Furthermore, traffic offloading among BSs can offer a network-level solution, wherein the cell-level traffic load can be dynamically adjusted to balance the energy supply and demand of BSs \cite{Tao_Magazine_RE} \cite{Zhongming}.
Although traffic offloading has been extensively investigated in on-grid cellular networks \cite{QYe_offloading}-\cite{EE_offloading_learning_JSAC}, the conventional offloading methods can not be applied when EH is leveraged. 
Instead, energy-aware traffic offloading schemes needs to be devised, i.e., the operations of each cell are optimized individually based on their renewable energy supply.
Energy-aware traffic offloading schemes have been proposed for single-tier homogeneous networks \cite{Tao_ICE_long} \cite{JGong_TC}, two-tier HCN with single HSBS \cite{BUPT_conf_Het_RE} and RSBS \cite{mine_GC_EH}, respectively.

Different from existing works, we focus on the design of energy-aware traffic offloading for HCNs with multiple SBSs powered by diverse energy sources.
We aim to minimize the on-grid network power consumption while satisfying the QoS requirement in terms of rate outage probability.
To this end, users are dynamically offloaded from the MBS to the SBSs, based on the statistical information of traffic intensity and renewable energy.
Accordingly, dynamic cell activation and power control are conducted at SBSs to provide on-demand service for energy saving.
To solve the optimization problem, the approximated outage probability is derived in closed form based on stochastic geometry, and the renewable energy supply and consumption are analyzed using M/D/1 queue.
For the single-SBS case, the power saving gain achieved by activating a CSBS, RSBS, or HSBS can be derived with respect to the amount of traffic offloaded, based on which the activation condition and the optimal amount of traffic offloaded are obtained.
Furthermore, for the multi-SBS case, mixed integer programming problems are formulated, and a two-stage energy-aware traffic offloading (TEATO) scheme is proposed accordingly.
In the first stage, the optimal amount of traffic offloaded from the MBS to each individual SBS is obtained based on the analytical results of the single-SBS cases.
In the second phase, the ON-OFF states of SBSs are optimized, which is further formulated as a 0-1 knapsack problem and solved by applying the Lagrange multiplier method.

The main contributions of this work are threefold:
\begin{enumerate}
  \item The power saving performance of traffic offloading is investigated. Specifically, the on-gird power saving gain achieved by offloading traffic from a conventional MBS to a RSBS, HSBS, or CSBS is derived in closed form, which reflects the conversion rate of harvested energy into on-grid power through traffic offloading.
  \item Based on the derived power saving gain, the optimal ON-OFF state of the SBS and corresponding traffic amount for offloading are determined, for the given statistical information of traffic demand and renewable energy arrival.
  \item A traffic offloading scheme is proposed for the HCNs with multiple SBSs, which can achieve significant on-gird power saving gain while satisfying the QoS requirement. Simulation results demonstrate that more than 50\% of the on-grid power consumption can be saved for typical daily traffic and solar energy profiles, compared with the conventional greedy traffic offloading methods where traffic is offloaded to the SBSs with priority without considering energy status or cell sleeping.
\end{enumerate}

The reminder of the paper is organized as follows. System model is introduced in Section~\ref{sec_system_model}. Section~\ref{sec_power_demand_supply} analyzes the power demand and supply for SBSs with EH devices. Then, the power consumption minimization problem for the single-SBS case is studied in Section~\ref{sec_Single_SBS}, and TEATO scheme is proposed for the multi-SBS case in Section~\ref{sec_Multiple_SBS}. Simulation results are presented in Section~\ref{sec_numerical_results}, followed by the conclusion in Section~\ref{sec_conclusions}.

\section{System Model}
    \label{sec_system_model}
    In this section, the details of the HCN with hybrid energy supply are presented as follows.
The key notations are also summarized in Table I.

\subsection{HCNs with Hybrid Energy Supply}

\begin{figure}
  \centering
  \includegraphics[width=3.0in]{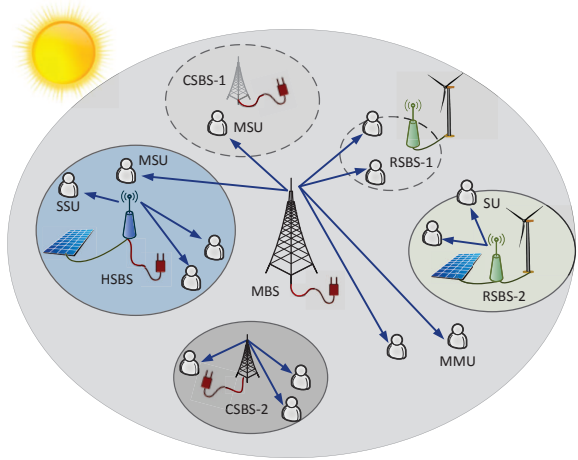}\\
  \caption{Illustration of a HCN with diverse energy sources.}\label{fig_HCN}
\end{figure}

With EH technology employed, a typical scenario of HCN is shown in Fig.~\ref{fig_HCN}, where different types of SBSs are deployed in addition to the conventional MBS to enhance network capacity.
Based on the energy source, SBSs can be classified into three types: (1) \textbf{CSBSs} powered by on-grid energy only; (2) \textbf{RSBSs} powered solely by harvested renewable energy (like solar and wind power); and (3) \textbf{HSBSs} powered jointly by energy harvesting devices and power grid.
Denote by $N_\mathrm{C}$, $N_\mathrm{R}$, and $N_\mathrm{H}$ the number of CSBSs, RSBSs, and HSBSs, respectively.
Denote by $\mathfrak{B}_\mathrm{C} = \{1,2,...,N_\mathrm{C}\}$, $\mathfrak{B}_\mathrm{R} = \{1,2,...,N_\mathrm{R}\}$, and $\mathfrak{B}_\mathrm{H} = \{1,2,...,N_\mathrm{H}\}$, the set of CSBSs, RSBSs, and HSBSs, respectively.
Let $\mathfrak{B} = \{\mathfrak{B}_\mathrm{C}, \mathfrak{B}_\mathrm{R}, \mathfrak{B}_\mathrm{H}\}$ be the set of all SBSs. 
SBS$_n$ serves a circular area with radius $D_{\mathrm{s},n}$, and the small cells are assumed to have no overlaps with each other.
The MBS is always active to guarantee the basic coverage, whereas SBSs can be dynamically activated for traffic offloading or deactivated for energy saving, depending on the traffic and energy status.
For example, in Fig.~\ref{fig_HCN}, the lightly-loaded CSBS-1 is deactivated to reduce the on-grid power consumption, while RSBS-1 is shut down due to the lack of harvested energy.

\begin{table}
\center
\caption{Notation Table}
\begin{tabular}{c|c}
  \hline
  \hline
  $D_0$ & coverage radius of the MBS\\  
  $D_{n}$ & coverage radius of SBS$_n$\\
  $\rho_0$ & user density outside of all small cells\\
  $\rho_{n}$ & user density in small cell $n$\\
  $\varphi_n$ & ratio of users offloaded to the SBS in small cell $n$\\
  $I_n$ & 0-1 indicator showing whether SBS$_n$ is active or not\\
  $E$ & a unit of energy\\
  $\lambda_{\mathrm{E},n}$ & arrival rate of per unit energy at SBS$_n$\\
  $\mu_{\mathrm{E},n}$ & consumption rate of per unit energy at SBS$_n$\\
  $R_\mathrm{Q}$ & data rate requirement of mobile users\\
  $\eta$ & required maximal outage probability\\
  $\sigma^2$ & noise power density\\
  $W_\mathrm{m}$ (/$W_\mathrm{s}$) & system bandwidth available for the MBS (/SBSs)\\
  $\theta_\mathrm{m}$ (/$\theta_\mathrm{s}$) & inter-cell interference among MBSs (/SBSs)\\
  $\alpha_\mathrm{m}$ (/$\alpha_\mathrm{s}$) & path loss factor of the MBS-tier (/SBS-tier)\\
  \hline
  \hline
\end{tabular}
\label{tab_P_BS}
\end{table}

\subsection{Traffic Model}

The user distribution in spatial domain is modeled as a non-homogeneous Poisson Point Process (PPP), whose density at time $t$ is $\rho_{n}(t)$ in small cell $n$ and $\rho_0(t)$ outside of all small cells.
As shown in Fig.~\ref{fig_HCN}, users located outside of the small cells can only be served by the MBS, while users within small cells can be partly or fully offloaded to the corresponding SBSs, according to the traffic and energy status\footnote{Under conventional offloading methods, all users within small cells should be offloaded to SBSs for better channel quality.}.
Thus, users can be classified into three types based on the serving BSs and location: (1) Macro-Macro Users (MMUs), users which are located outside of small cells and served by the MBS; (2) SBS-SBS Users (SSUs), users located within small cells and offloaded to SBSs; (3) Macro-SBS Users (MSUs), users located in small cells but served by the MBS.
For theoretical analysis, random offloading scheme is adopted, where users in small cell $n$ are offloaded to SBS$_n$ with probability $\varphi_n(t)$ and are served by the MBS with probability $1-\varphi_n(t)$.
Thus, the distributions of SSUs and MSUs in small cell $n$ also follow PPP with density $\varphi_n(t) \rho_{n}(t)$ and $(1-\varphi_n(t)) \rho_{n}(t)$ respectively, according to the properties of PPP \cite{Stochastic_Geometry}.
Note that we focus on large time-scale operation, during which the instantaneous time-varying channel quality can be ignored for offloading decisions.
Besides, the random scheme can work as a benchmark, which has been widely adopted for network performance analysis.

As for spectrum resource, the bandwidths available to the MBS-tier and SBS-tier are orthogonal to avoid cross-tier interference, whereas the intra-tier spectrum reuse factor is set as 1.
Denote by $W_\mathrm{m}$ and $W_\mathrm{s}$ the system bandwidth available to the MBS and each SBS, respectively.
For each BS, its available bandwidth can be partially deactivated to reduce the power consumption, i.e., power control.
At the SBS$_n$, the bandwidth actually utilized is denoted as $w_{\mathrm{ss},n}\leq W_\mathrm{s}$, which is allocated to its SSUs equally for fairness.
At the MBS, $W_\mathrm{m}$ is further divided into different orthogonal portions: $w_\mathrm{mm}$ for serving MMUs and $w_{\mathrm{ms},n}$ for serving MSUs in small cell $n$, where $w_\mathrm{mm} + \sum_{n=1}^{N_\mathrm{C}+N_\mathrm{R}+N_\mathrm{H}} w_{\mathrm{ms},n} \leq W_\mathrm{m}$.
In addition, $w_{\mathrm{ss},n}$, $w_\mathrm{mm}$ and $w_{\mathrm{ms},n}$ should be dynamically adjusted to satisfy the QoS requirements of SSUs, MMUs and MSUs.

\subsection{Power Consumption Model}

\begin{table}
\center
\caption{Power model parameters for different types of BSs}
\begin{tabular}{cccc}
  \hline
  \hline
  & {\begin{tabular}{c}Transmit Power\\$P_\mathrm{T}$ (W)\end{tabular}} & {\begin{tabular}{c}Constant Power\\$P_\mathrm{C}$ (W)\end{tabular}} & {\begin{tabular}{c}Coefficient\\$\beta$ \end{tabular}} \\
  \hline
  Macro & 20 & 130 & 4.7 \\
  Micro & 6.3 & 56 & 2.6 \\
  Pico & 0.13 & 6.8 & 4 \\
  Femto & 0.05 & 4.8 & 8 \\
  \hline
  \hline
\end{tabular}
\label{tab_P_BS}
\end{table}

BSs can work in either \emph{active mode} or \emph{sleep mode}, with different power consumption parameters.
According to the EARTH project, the power consumption of a BS in \emph{active mode} can be modeled as a constant power term plus a radio frequency (RF) related power \cite{EARTH}:
\begin{equation}\label{eq_P_BS}
    P_\mathrm{BS} = P_\mathrm{C} + \beta P_\mathrm{RF},
\end{equation}
where $P_\mathrm{C}$ denotes an offset of site power including the baseband processor, the cooling system and etc., coefficient $\beta$ is the inverse of power amplifier efficiency factor, and $P_\mathrm{RF}$ is the RF power.
The power related parameters for different types of BSs are given in Table~\ref{tab_P_BS} \cite{EARTH}.

The system bandwidth is further divided into orthogonal subcarriers, and the BS can decide how many subcarriers are utilized depending on the traffic demand.
The RF power is proportional to the bandwidth of utilized subcarriers $w$, i.e.,
\begin{equation}\label{eq_P_RF}
    P_\mathrm{RF} = \frac{w}{W} P_\mathrm{T},~~0 < w \leq W,
\end{equation}
where $W$ is the available system bandwidth and $P_\mathrm{T}$ is the transmit power level.
In this work, a constant power level is considered according to the LTE standard \cite{LTE_standard}, i.e., $P_\mathrm{T}$ is treated as a system parameter, while we control the RF power by adjusting the utilized bandwidth $w$.
Substituting $P_\mathrm{RF}$ in (\ref{eq_P_BS}) with (\ref{eq_P_RF}), we have
\begin{equation}\label{eq_P_BS_2}
    P_\mathrm{BS} = P_\mathrm{C} + \frac{w}{W} \beta P_\mathrm{T}.
\end{equation}

If a BS is completely turned off (switch to \emph{sleep mode}), a small amount of power is still consumed so that the BS can be reactivated.
Considering that the power needed is negligible compared with $P_\mathrm{C}$, the power consumption in sleep mode can be approximated as zero \cite{LTE_standard}.

\subsection{Green Energy Supply Model}

\begin{figure}
  \centering
  \includegraphics[width=3.0in]{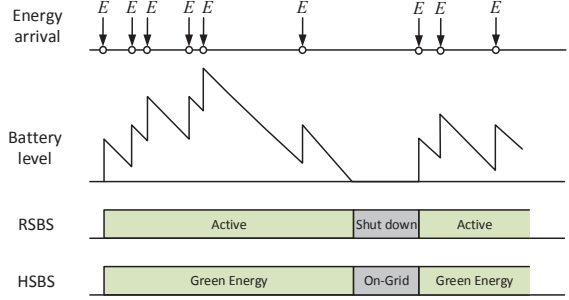}\\
  \caption{Renewable energy arrival and consumption process.}\label{fig_work_mode}
\end{figure}

Discrete energy model is adopted to describe the process of energy harvesting, and a unit of energy is denoted by $E$ \cite{xiaoxia_EH_d2d}.
Denote by $\lambda_{\mathrm{E},n}(t)$ the arrival rate of per unit energy at SBS$_n$ and time $t$.
The harvested energy is saved in its battery for future use.
The battery is considered to have sufficient capacity for realistic operation conditions, and thus we assume no battery overflow happens.

Fig.~\ref{fig_work_mode} shows an example of the energy supply and consumption process at a typical SBS, whereby the harvested energy is used to power the SBS whenever the battery is not empty.
For RSBSs without grid power input, they have to be shut down when the battery runs out.
Consequently, the corresponding users will be served by the upper-tier MBS for QoS guarantee.
Note that handover procedure is conducted when the RSBS is shut down or reactivated, causing additional signaling overhead and power consumption.
For HSBSs, they can use the backup energy (i.e., on-grid power) when there is no green energy, until renewable energy arrives.

The energy supply and consumption process of each SBS can be modeled by a queue, where the queue length denotes the battery amount \cite{EH_single_link_2}.
Based on the power consumption model of BSs (Eq.~(\ref{eq_P_BS_2})),  the equivalent service rate of per unit energy for SBS$_n$ is given by
\begin{equation}\label{eq_mu}
    \mu_{\mathrm{E},n}(t) = \frac{1}{E} \left(P_{\mathrm{Cs},n} + \frac{w_{\mathrm{ss},n}(t)}{W_\mathrm{s}}\beta_{\mathrm{s},n} P_{\mathrm{Ts},n} \right),
\end{equation}
where $P_{\mathrm{Ts},n}$, $P_{\mathrm{Cs},n}$ and $\beta_{\mathrm{s},n}$ are the transmit power, constant power and power amplifier coefficient of SBS$_n$, respectively.
$\mu_{\mathrm{E}}$ is called \emph{Energy Consumption Rate} in the rest of this paper for simplicity.
Notice that the energy consumption rate can be adjusted by changing the utilized bandwidth, which affects the traffic service capability of SBS$_n$ on the other hand.

	\subsection{Wireless Communication Model}

        If user $u$ is served by SBS$_n$, its received SINR is given by \cite{Jeffery_fundamental}
            \begin{equation}\label{eq_SINR_SBS}
                \gamma_{\mathrm{ss},nu} = I_n \frac{P_{\mathrm{Ts},n} w_u}{W_\mathrm{s}} \frac{ {d_{nu}}^{-\alpha_\mathrm{s}} h_{nu} } { \left( \theta_\mathrm{s} +1 \right) \sigma^2 w_u },
            \end{equation}
        where $I_n$ is a 0-1 indicator denoting whether SBS$_n$ is active or not, $w_u$ is the bandwidth allocated to user $u$, $d_{nu}$ is the distance between user $u$ and SBS$_n$, $\alpha_\mathrm{s}$ is the path loss exponent of the SBS-tier, $h_{nu}$ is an exponential random variable with unit mean reflecting the effect of Rayleigh fading, $\theta_\mathrm{s}$ is the ratio of inter-cell interference to noise among SBSs, and $\sigma^2$ is the noise power density.
        As each SBS allocates bandwidth equally to its associated users, the achievable rate of a generic user $u$ is as follows:
            \begin{equation}\label{eq_r_s}
                r_{\mathrm{ss},nu} = \frac{w_{\mathrm{ss},n}}{K_{\mathrm{ss},n}+1} \log_2 (1 + \gamma_{\mathrm{ss},nu}),
            \end{equation}
        where the random variable $K_{\mathrm{ss},n}$ denotes the number of residential SSUs of SBS$_n$ except user $u$.

        Similarly, if user $u$ is served by the MBS as a MMU or MSU, its received SINR is given by
            \begin{equation}\label{eq_SINR_SBS}
                \gamma_{\mathrm{m},u} = \frac{P_{\mathrm{Tm}} w_u}{W_\mathrm{m}} \frac{ {d_{0u}}^{-\alpha_\mathrm{m}} h_{0u}  } { (\theta_\mathrm{m}+1) \sigma^2 w_u },
            \end{equation}
        where $P_\mathrm{Tm}$ is the transmit power level of the MBS, $d_{0u}$ is the distance between user $u$ and the MBS, $\alpha_\mathrm{m}$ is the path loss exponent of the MBS-tier, $\theta_\mathrm{m}$ is the interference to noise ratio from other MBSs, and $h_{0u}$ reflects Rayleigh fading with the same probability distribution as $h_{nu}$.
        Then, the achievable rate of user $u$ is given by
            \begin{equation}\label{eq_r_s}
                \begin{split}
                r_{\mathrm{mm},u} & = \frac{w_{\mathrm{mm}}}{K_{\mathrm{mm}}+1} \log_2 (1 + \gamma_{\mathrm{m},u}), ~~\mbox{for MMU},\\
                r_{\mathrm{ms},nu} & = \frac{w_{\mathrm{ms},n}}{K_{\mathrm{ms},n}+1} \log_2 (1 + \gamma_{\mathrm{m},u}), ~~\mbox{for MSU},
                \end{split}
            \end{equation}
        where $K_\mathrm{mm}$ and $K_{\mathrm{ms},n}$ denote the number of residual MMUs and MSUs, respectively.

\section{Analysis of Power Supply and Demand}
    \label{sec_power_demand_supply}
    Two time scales are considered for the problem analysis.
In the large time scale, we divide the time into $T$ periods (e.g., $T=24$ and the length of each time period is 1 hour), and assume the average energy harvesting rate and user density remain static in each time period, but may change over different periods.
During period $t$, the arrival of renewable energy packets is modeled as Poisson process with rate $\lambda_{\mathrm{E},n}(t)$ for SBS$_n$, and the distribution of user in small cell $n$ follows PPP with density $\rho_n(t)$.
Notice that the battery level, the location and the number of users vary randomly in the small time scale (e.g., tens of milliseconds).

In this work, we optimize the amount of offloaded traffic, the ON-OFF state, and the RF power of each SBS at the large time scale, based on the stochastic information of traffic and energy (i.e., energy arrival rate and user density). 
The optimization is conducted for each period independently, and thus we can focus on the optimization for one period\footnote{The subscript $t$ is omitted in the following to ease the presentation.}.

To start with, we analyze the green power supply and demand in this section.
Specifically, the battery level is analyzed with queueing theory and the QoS performance (i.e., outage probability) is derived with stochastic geometry, considering the small-scale randomness.

\subsection{Energy Queue Analysis}
For a SBS with EH, the variation of battery can be modeled as a M/D/1 queue with arrival rate $\lambda_\mathrm{E}$ and service rate $\mu_\mathrm{E}$ given by Eq.~(\ref{eq_mu}).
In what follows, we analyze the stable status of the energy queue.
For the M/D/1 queue, the embedded Markov chain method is usually applied to analyze the stable status \cite{MD1_queue}.
Denote $L$ the queue length at any time and $L^+$ the queue length when a unit energy leaves the energy queuing.
As the energy arrival is Poisson process, the transition of states $L^+$ is memoryless.
Thus, the state $L^+$ can be modeled as a Markov chain, with the transition probability matrix given by
\begin{equation}\label{eq_MD1_transition_matrix}
     A = \left( \begin{array}{ccccc} a_0 & a_1 & a_2 & a_3 & \cdots \\ a_0 & a_1 & a_2 & a_3 & \cdots \\ 0 & a_0 & a_1 & a_2 & \cdots \\ 0 & 0& a_0 & a_1 & \cdots \\ \vdots & \vdots & \vdots & \vdots & \ddots \end{array} \right) ,
\end{equation}
where
\begin{equation}\label{eq_MD1_a}
    a_i = \frac{1}{i!} \left(\frac{\lambda_\mathrm{E}}{\mu_\mathrm{E}}\right)^{-i} e^{-\frac{\lambda_\mathrm{E}}{\mu_\mathrm{E}}}, i = 0,1,\cdots .
\end{equation}

When $\frac{\lambda_\mathrm{E}}{\mu_\mathrm{E}} \geq 1$, the queue is not stable and the queue length goes to infinity, which means that the harvested energy is always sufficient.
When $\frac{\lambda_\mathrm{E}}{\mu_\mathrm{E}}<1$, the stationary probability distribution of $L^+$ can be derived by Pollaczek-Khinchin formula \cite{MD1_queue}, i.e.,
\begin{equation}\label{eq_MD1_steady}
    \begin{split}
        q_0 = & 1-\frac{\lambda_\mathrm{E}}{\mu_\mathrm{E}} \\
        q_1 = & (1-\frac{\lambda_\mathrm{E}}{\mu_\mathrm{E}})(e^{\frac{\lambda_\mathrm{E}}{\mu_\mathrm{E}}} - 1)\\
        q_{L^+} = & (1-\frac{\lambda_\mathrm{E}}{\mu_\mathrm{E}})\left\{ e^{\frac{\lambda_\mathrm{E}}{\mu_\mathrm{E}}{L^+} } + \sum_{k=1}^{{L^+}-1}e^{k \frac{\lambda_\mathrm{E}}{\mu_\mathrm{E}}} (-1)^{\frac{\lambda_\mathrm{E}}{\mu_\mathrm{E}} -k} \right.\\
        & \left. \cdot \left[\frac{(k \frac{\lambda_\mathrm{E}}{\mu_\mathrm{E}})^{{L^+}-k}}{({L^+}-k)!}+\frac{(k \frac{\lambda_\mathrm{E}}{\mu_\mathrm{E}} )^{b-{L^+}-1}}{(b-{L^+}-1)!} \right]\right\} \quad ({L^+}>2),
    \end{split}
\end{equation}
For the M/D/1 queue, we have $q_L = q_{L^+}$ \cite{MD1_queue}.
Thus, the stationary probability distribution of the energy queue length (i.e., the amount of available green energy) is derived.

\subsection{Outage Probability Analysis}

Service outage happens when the user's achievable data rate is less than the requirement $R_\mathrm{Q}$, due to channel fading or bandwidth limitation.
The outage probability should be guaranteed to be below a certain threshold $\eta$.
We are interested in analyzing the outage probability constraint for SSUs, MMUs and MSUs, respectively, based on which the power demand can be obtained.

According to the wireless communication model, the outage probability of a typical SSU $u$ of SBS$_n$ is given by
        \begin{equation}\label{eq_outage_SBS}\footnotesize
            \begin{split}
            G_{\mathrm{ss},n} & = \mathbb{E}_{\{K_{\mathrm{ss},n}, d_{\mathrm{ss},n}\}} \left\{ \mathds{P} \left( r_{\mathrm{ss},nu} < R_\mathrm{Q} \Big| K_{\mathrm{ss},n}, d_{\mathrm{ss},n} \right) \right\}\\
            & = \int\limits_0\limits^{D_{\mathrm{s},n}} \sum\limits_{k=0}\limits^{\infty} \mathds{P}\left\{ \gamma_{\mathrm{ss},n} < 2^{\frac{(k+1)R_\mathrm{Q}}{w_{\mathrm{ss},n}}} - 1 | d \right\} Q_{K_{\mathrm{ss},n}}(k) f_{d_{\mathrm{ss},n}}(d) \mathrm{d} d ,
            \end{split}
        \end{equation}
where $Q_{K_{\mathrm{ss},n}}(k)$ is the probability that SBS$_n$ serves $k$ residential SSUs except SSU $u$, and $f_{d_{\mathrm{ss},n}}(d) = \frac{2 \pi}{D_{n}} d$ is the probability density function (PDF) of the distance between $u$ and SBS$_n$.
As the distribution of users follows PPP in each small cell, $K_{\mathrm{ss},n}$ follows the Poisson distribution of parameter $\pi \varphi \rho_{n} {D_{n}}^2$ according to Slivnyak-Mecke theorem \cite{Slivnyak_theorem}, where $\varphi$ is the offloading ratio.
Although the outage probability of Eq.~(\ref{eq_outage_SBS}) cannot be derived in general case, the closed-form expression can be obtained in the region of high SINR and large system bandwidth, given as Theorem~1.\\

    \textbf{Theorem 1.} As $\frac{P_{\mathrm{Ts},n}} {(\theta_\mathrm{s}+1)\sigma^2 W_\mathrm{s}}\rightarrow \infty$ (i.e., SBS$_n$ provides high SINR) and $\frac{R_\mathrm{Q}} {w_{\mathrm{ss},n}} \rightarrow 0$ (i.e., sufficient system bandwidth), the service outage probability is given as follows:
            \begin{equation}\label{eq_outage_SBS_closed_2}\footnotesize
                G_{\mathrm{ss},n} \!=\! \frac{2 {D_{n}}^{\alpha_\mathrm{s}} (\theta_\mathrm{s}\!+\!1)\sigma^2 }{  \left(\alpha_\mathrm{s}\!+\!2\right) P_{\mathrm{Ts},n} {W_\mathrm{s}}^{-1}} \! \left(2^{\frac{R_\mathrm{Q}}{w_{\mathrm{ss},n}}\left(1 + \pi {D_{n}}^2 \varphi_n \rho_{n} \frac{R_\mathrm{Q}}{w_{\mathrm{ss},n}} \right)}\!-\!1\right).
            \end{equation}
    \emph{Proof}: Please refer to Appendix~\ref{appendix_SBS}.
    \hfill \rule{4pt}{8pt}

In fact, the assumptions of Theorem~1 are reasonable in practical cellular systems, where the SINR of users is generally high enough for reliable communications.
Besides, each BS can support a large number of users simultaneously, which requires large amount of system bandwidth.
Thus, the system bandwidth should be much higher than the data rate requirement. \
Therefore, Eq.~(\ref{eq_outage_SBS_closed_2}) can be applied to approximate the outage probability for problem analysis.

Notice that the outage probability Eq.~(\ref{eq_outage_SBS_closed_2}) constraints the required bandwidth for given traffic density. 
Specifically, the service outage constraint of SSUs $G_{\mathrm{ss},n} \leq \eta$ can be written as
        \begin{equation}\label{eq_outage_SBS_simple}
            \bar{w}_{\mathrm{ss},n} \tau_{\mathrm{ss},n} \geq R_\mathrm{Q},
        \end{equation}
where $\bar{w}_{\mathrm{ss},n}=\frac{w_{\mathrm{ss},n}}{1+ \varphi_n \rho_{n} \pi {D_{n}}^2}$ is the expected bandwidth allocated to each SSU, and $\tau_{\mathrm{ss},n}$ denotes the spectrum efficiency of cell edge users given by
        \begin{equation}\label{eq_tau_SBS}
            \tau_{\mathrm{ss},n} = \log_2 \left( 1+ \frac{P_{\mathrm{Ts},n}}{(\theta_\mathrm{s}+1)} \frac{\alpha_\mathrm{s}+2}{2 \sigma^2 W_\mathrm{s}} \frac{\eta}{{D_{n}}^{\alpha_\mathrm{s}}} \right).
        \end{equation}
The physical meaning of Eq.~(\ref{eq_outage_SBS_simple}) is that the average data rate of the non-cell-edge users (with spectrum efficiency above $\tau_{\mathrm{ss},n}$) should be no smaller than $R_\mathrm{Q}$.

The outage probability of MMUs cannot be derived in closed form, which varies with the location and coverage of each SBS.
To obtain the analytical result, we approximate the MMUs to be uniformly distributed in the macro cell with density $\rho'_\mathrm{0}$ as follows:
\begin{equation}\label{eq_rho_m_uniform}
    \rho'_\mathrm{0} = \frac{\rho_\mathrm{0}}{{D_\mathrm{0}}^2} \left( {D_\mathrm{0}}^2 - \sum\limits_{n\in\mathfrak{B}} D_{n}^2 \right).
\end{equation}
Thus the approximated outage probability of MMUs can be derived in the same way as Theorem~1, and the constraint $G_\mathrm{mm}\leq \eta$ is equivalent to
            \begin{equation}\label{eq_outage_MBS_simple}
                \bar{w}_\mathrm{mm} \tau_\mathrm{mm} \geq R_\mathrm{Q},
            \end{equation}
where $\bar{w}_\mathrm{mm}=\frac{w_\mathrm{mm}}{ 1+ \pi {D_\mathrm{0}}^2 \rho'_\mathrm{0} }$ and $\tau_\mathrm{mm}$ are given by
            \begin{equation}\label{eq_tau_MBS}
                \tau_\mathrm{mm} = \log_2 \left( 1+ \frac{P_\mathrm{Tm}}{(\theta_\mathrm{m}+1)} \frac{\alpha_\mathrm{m}+2}{2 \sigma^2 W_\mathrm{m}} \frac{\eta}{{D_\mathrm{0}}^{\alpha_\mathrm{m}}} \right).
            \end{equation}
Notice that the approximation is reasonable when the SBSs are considered to be uniformly distributed in the macro cell.

Next, we consider an MSU $u$ served by SBS$_n$.
By approximating $u$ located at SBS$_n$, the closed-form outage probability is given by Theorem~2.

    \textbf{Theorem~2.} As $\frac{P_\mathrm{Tm}}{(\theta_\mathrm{m}+1)\sigma^2 W_\mathrm{m}}\rightarrow \infty$ (i.e., the MBS provides high SINR) and $\frac{R_\mathrm{Q}}{w_{\mathrm{ms},n}}\rightarrow 0$ (i.e., sufficient system bandwidth), the outage probability constraint of the MSUs $G_{\mathrm{ms},n} \leq \eta$ can be approximated as
            \begin{equation}\label{eq_outage_ms_simple}
                \bar{w}_{\mathrm{ms},n} \tau_{\mathrm{ms},n} \geq R_\mathrm{Q},
            \end{equation}
    where
            \begin{equation}
                \begin{split}
                \bar{w}_{\mathrm{ms},n} &=\frac{w_{\mathrm{ms},n}}{1+ (1-\varphi_n) \rho_{n} \pi {D_{n}}^2}, \\ 
                \tau_{\mathrm{ms},n}\!  &=\! \log_2 \left( 1\!+ \!\frac{\eta P_\mathrm{Tm}}{\sigma^2 W_\mathrm{m} (\theta_\mathrm{m}+1) {D_{\mathrm{ms},n}}^{\alpha_\mathrm{m}}} \right)
                \end{split}
            \end{equation}
    and $D_{\mathrm{ms},n}$ denotes the distance between the MBS and SBS$_n$.

    \hfill \rule{4pt}{8pt}

Theorem~2 can be proved in the similar way as Theorem~1, and thus the detailed proof is omitted due to space limitations.
When the SBS$_n$ is in sleep mode, the service outage probability of users in the small cell coverage can also be obtained based on Theorem~2, by setting $\varphi_n=0$, i.e., no traffic is offloaded to SBS$_n$.

Notice that Eqs.~(\ref{eq_outage_SBS_simple}), (\ref{eq_outage_MBS_simple}), and (\ref{eq_outage_ms_simple}) constrain the minimum bandwidth required for the given traffic demand ($\rho_\mathrm{0}$, $\rho_{n}$) and the offloading scheme $\varphi_n$, which can reflect the power demand of each SBS and MBS according to the BS power consumption model Eq.~(\ref{eq_P_BS_2}).

\section{Power Consumption Minimization for Single-SBS Case}
    \label{sec_Single_SBS}
    In this section, we optimize the traffic offloading for the single small cell case, where the HSBS and RSBS are analyzed, respectively.
Note that a CSBS can be considered as a HSBS whose energy arrival rate is set to zero, i.e., $\lambda_\mathrm{E}=0$.

\subsection{Single-HSBS Case}

For the single-HSBS case, the total on-grid power consumption $P_{\mathrm{sum}}$ consists of two parts:
\begin{equation}\label{eq_P_total_HSBS}
    P_{\mathrm{sum}} = P_\mathrm{MBS} + P_\mathrm{HSBS},
\end{equation}
where $P_\mathrm{MBS}$ and $P_\mathrm{HSBS}$ denote the on-grid power consumptions of the MBS and HSBS, respectively.
$P_\mathrm{MBS}$ and $P_\mathrm{HSBS}$ can be derived based on Eq.~(\ref{eq_P_BS_2}).
Denote $I_\mathrm{H}$ a 0-1 variable indicating whether the HSBS is active or not, while $w_\mathrm{ms}^{(\mathrm{a})}$ and $w_\mathrm{ms}^{(\mathrm{o})}$ the corresponding bandwidth needed by the MSUs.
Then we have
\begin{equation}\label{eq_sum_power_HSBS_detail} \small
    P_\mathrm{MBS} = P_{\mathrm{Cm}} \!+\! \frac{\beta_\mathrm{m}P_\mathrm{Tm}}{W_\mathrm{m}} \! \left( w_\mathrm{mm} \!+\! I_\mathrm{H} w_\mathrm{ms}^{(\mathrm{a})} \!+\! (1\!-\!I_\mathrm{H}) \! w_\mathrm{ms}^{(\mathrm{o})} \right).
\end{equation}
In addition, as the HSBS consumes on-grid power only when the battery is empty, we have
\begin{equation}\label{eq_sum_power_HSBS_detail_2}
    P_{\mathrm{HSBS}} = I_\mathrm{H} q_0 \left(P_{\mathrm{Cs}} + \frac{\beta_\mathrm{s}P_\mathrm{Ts}}{W_\mathrm{s}} w_\mathrm{ss} \right),
\end{equation}
where $q_0$ is the probability of empty energy queue, obtained from  Eq.~(\ref{eq_MD1_steady}).

Then, the on-grid power consumption minimization problem of the single-HSBS case can be formulated as follows:
\begin{subequations}\label{eq_Problem_HSBS}
    \begin{align}
    \mathcal{P}1:~~~&\min\limits_{I_\mathrm{H},\mu_\mathrm{E}} ~~P_{\mathrm{sum}} \\
    \mbox{s.t.}~~~ & G_\mathrm{mm}\leq \eta, G_\mathrm{ss}\leq \eta, G_\mathrm{ms} \leq \eta, \\
                & 0 \!\leq \! w_\mathrm{mm} \!+\! I_\mathrm{H} w_\mathrm{ms}^{(\mathrm{a})} \!+\!  (1-I_\mathrm{H}) w_\mathrm{ms}^{(\mathrm{o})} \!\leq\! W_\mathrm{m}, \\
                & 0\leq w_\mathrm{ss} \leq W_\mathrm{s},
    \end{align}
\end{subequations}
where Eq.~(\ref{eq_Problem_HSBS}b) guarantees the QoS, Eq.~(\ref{eq_Problem_HSBS}c) and  Eq.~(\ref{eq_Problem_HSBS}d) are due to the bandwidth limitations of MBS and HSBS, respectively. $\mu_\mathrm{E}$ can be derived based on the power consumption model of HSBS Eq.~(\ref{eq_mu}).

Intuitively, there exists a tradeoff between the power consumptions of the MBS and HSBS.
By activating the HSBS for traffic offloading, the traffic load of the MBS decreases, reducing the RF power consumption of the MBS.
Whereas, the activated HSBS introduces additional on-grid power consumption, especially when the renewable energy is insufficient.
Furthermore, higher energy consumption rate $\mu_\mathrm{E}$ means more users offloaded from the MBS to the HSBS, which reduces RF power of the MBS but increases the power demand at the HSBS.
In what follows, we analyze this tradeoff relationship to solve problem $\mathcal{P}1$.

Denote by $\Delta_\mathrm{H}$ the power saving gain by activating the HSBS:
\begin{equation}\label{eq_P_gain_HSBS_definition}
    \Delta_\mathrm{H} = \left\{P_{\mathrm{sum}}|I_\mathrm{H}=0\right\} - \left\{P_{\mathrm{sum}}|I_\mathrm{H}=1\right\}.
\end{equation}
Based on the results of Theorems~1 and 2, the closed-form expression of $\Delta_\mathrm{H}$ is given as Theorem~3.
Then, the optimal solution of problem $\mathcal{P}1$ is derived as Theorem~4.

\textbf{Theorem~3.} The power saving gain by activating a HSBS for traffic offloading is given as follows:
\begin{equation}\label{eq_gain_hsbs_2} \footnotesize
    \Delta_\mathrm{H} \!=\! \left\{ \begin{array}{ll} \zeta_\mathrm{EE} \mu_\mathrm{E} E - \zeta_\mathrm{EE} P_{\mathrm{Cs}} - \frac{R_\mathrm{Q}\beta_\mathrm{m}P_\mathrm{Tm}}{\tau_\mathrm{ms} W_\mathrm{m}}, &\mu_\mathrm{E}\leq\lambda_\mathrm{E}\\
    \vspace{0.05in}\\
    \left[ \zeta_\mathrm{EE} \!-\!1 \right] \mu_\mathrm{E} E \!-\! \zeta_\mathrm{EE}  P_{\mathrm{Cs}} \!-\! \frac{R_\mathrm{Q} \beta_\mathrm{m}P_\mathrm{Tm}}{\tau_\mathrm{ms} W_\mathrm{m}} \!+\! \lambda_\mathrm{E} E,  & \mu_\mathrm{E}>\lambda_\mathrm{E} \end{array} \right. ,
\end{equation}
where $\zeta_\mathrm{EE} = \frac{W_\mathrm{s}\tau_\mathrm{ss} \beta_\mathrm{m} P_\mathrm{Tm}} {W_\mathrm{m} \tau_\mathrm{ms} \beta_\mathrm{s} P_\mathrm{Ts}}$.

\emph{Proof}: Please refer to Appendix~\ref{appendix_HSBS}. \hfill \rule{4pt}{8pt}

\textbf{Theorem~4.} If the HSBS is active, the optimal energy consumption rate satisfies
\begin{equation}\label{eq_mu_opt_HSBS} \small
    \tilde{\mu}_\mathrm{E} = \frac{1}{E} \left\{ P_{\mathrm{Cs}}+ \min \left\{1, \frac{R_\mathrm{Q}}{\tau_\mathrm{ss} W_\mathrm{s}}\left(\rho_\mathrm{s}\pi {D_\mathrm{s}}^2 +1 \right) \right\} \beta_\mathrm{s} P_\mathrm{Ts} \right\},
\end{equation}
where $\rho_\mathrm{s}$ is the user density in the small cell, $D_\mathrm{s}$ and $P_\mathrm{Ts}$ denote the coverage radius and transmit power of the SBS.
In addition, the HSBS should be activated (i.e., $\tilde{I}_\mathrm{H}=1$) if $\Delta_\mathrm{H}|_{{\mu}_\mathrm{E}=\tilde{\mu}_\mathrm{E}}>0$ or $w_\mathrm{mm}+w_\mathrm{ms}^{(\mathrm{o})}>W_\mathrm{m}$; otherwise, $\tilde{I}_\mathrm{H}=0$.
\hfill \rule{4pt}{8pt}

\emph{Proof}: Please refer to Appendix~\ref{appendix_HSBS}. \hfill \rule{4pt}{8pt}

Notice that Theorem~3 reflects the conversion rate of harvested energy (i.e., $\lambda_\mathrm{E}$) into on-grid power (i.e., $\Delta_\mathrm{H}$), i.e., how much on-grid power can be saved with per unit harvested power.
In addition, the physical meaning of $\frac{W_\mathrm{s}\tau_\mathrm{ss}} {\beta_\mathrm{s} P_\mathrm{Ts}}$ is the average energy efficiency of the SSUs in bit/J, i.e., the of amount information transmitted with 1 J transmit power at the HSBS.
Similarly, $\frac{W_\mathrm{m} \tau_\mathrm{ms}}{\beta_\mathrm{m} P_\mathrm{Tm}} $ is the average energy efficiency of MSUs at the MBS.
Thus, $\zeta_\mathrm{EE}$ compares the energy efficiency of the HSBS and MBS.
Therefore, Theorem~3 indicates that users should be served by the BS (MBS or HSBS) which provides higher energy efficiency, if the harvested energy is insufficient to support the HSBS (i.e., $\mu_\mathrm{E}>\lambda_\mathrm{E}$).
In practice, the HSBS usually provides higher energy efficiency compared with the MBS, due to shorter transmission distance and lower path loss.
As a result, more subcarriers should be utilized to offload more users if the HSBS is active, which explains Theorem~4.
As for cell activation, $\Delta_\mathrm{H}|_{{\mu}_\mathrm{E}=\tilde{\mu}_\mathrm{E}}>0$ indicates activating the HSBS brings positive power saving gain.
Besides, $w_\mathrm{mm}+w_\mathrm{ms}^{(\mathrm{o})}>W_\mathrm{m}$ happens when the MBS is overloaded, in which case the HSBS should be activated to relieve the burden of the MBS.
For better understanding, typical asymptotic cases are illustrated in Corollaries~1 and 2.

\textbf{Corollary~1.} When $\lambda_\mathrm{E}\rightarrow 0$ and $\rho_\mathrm{s}\rightarrow 0$, activating the HSBS does not reduce the RF power of MBS but consumes on-grid energy $P_{\mathrm{Cs}}$. Thus, the HSBS should be deactivated.

\textbf{Corollary~2.} When $\lambda_\mathrm{E} \geq \tilde{\mu}_\mathrm{E}$ (i.e., sufficient green energy supply), offloading users to SBS reduces on-gird power consumption (i.e., $\Delta_\mathrm{H}|_{{\mu}_\mathrm{E}=\tilde{\mu}_\mathrm{E}} \geq 0$) and therefore the HSBS should be active.

\subsection{Single-RSBS Case}

Unlike the HSBS, the RSBS does not consume on-grid power.
Whereas, the SSUs have to be served by the MBS when the battery is empty, which causes handover, additional signaling cost and on-grid power consumption.
The average power consumption is given by
\begin{equation}\label{eq_sum_P_RSBS}
    P_\mathrm{sum} = P_\mathrm{MBS} +  P_\mathrm{HO},
\end{equation}
where $P_\mathrm{MBS}$ is the power consumption of the MBS, and $P_\mathrm{HO}$ reflects the additional power consumed by SSU handover.

Denote by $I_\mathrm{R} \in \{0,1\}$ the ON-OFF state of the RSBS.
If the RSBS is active, handover happens in the following cases: (1) RSBS is shut down when the battery runs out; (2) RSBS is reactivated when new energy arrives.
According to the energy queueing model, the first case corresponds to the event when $L^+$ transits from 1 to 0, with frequency of $q_1 A_{21} \mu_\mathrm{E}$ after the energy queue becomes stable.
Due to the duality between the two cases, the additional handover power consumption is given by
\begin{equation}\label{eq_RSBS_switching}
    \begin{split}
    P_\mathrm{HO} & =  I_\mathrm{R} \cdot 2 q_1 A_{21} \mu_\mathrm{E} C_\mathrm{HO}, \\
    & = \left\{ \begin{array}{ll}  2 I_\mathrm{R} (1-\frac{\lambda_\mathrm{E}}{\mu_\mathrm{E}})\left(1-e^{-\frac{\lambda_\mathrm{E}}{\mu_\mathrm{E}}}\right) \mu_\mathrm{E} C_\mathrm{HO}, & \lambda_\mathrm{E} < \mu_\mathrm{E} \\
    0, & \lambda_\mathrm{E} \geq \mu_\mathrm{E} \end{array} \right.
    \end{split}
\end{equation}
where $C_\mathrm{HO}$ denotes the energy consumed by one handover process in Joule.

Note that the SBS may be shut down due to energy shortage even when its state is set as \emph{on}, in which the MBS has to utilize more bandwidth to serve the SSUs with additional bandwidth.
Based on Eq.~(\ref{eq_P_BS_2}), the average on-grid power consumption of the MBS is given as follows:
\begin{equation}\label{eq_P_mbs_RSBS_on}
	\begin{split}
	    P_\mathrm{MBS} = &  P_{\mathrm{Cm}} + \frac{\beta_\mathrm{m}P_\mathrm{Tm}}{W_\mathrm{m}} \left(w_\mathrm{mm} + I_\mathrm{R} \left( (1-q_0) w_\mathrm{ms}^{(\mathrm{a})}  \right. \right.\\
	    & \left. \left. + q_0 w_\mathrm{ms}^{(\mathrm{o})} \right)  + (1-I_\mathrm{R}) w_\mathrm{ms}^{(\mathrm{o})} \right),
    \end{split}
\end{equation}
where $w_\mathrm{mm}$ is constrained by Eq.~(\ref{eq_outage_MBS_simple}), $w_\mathrm{ms}^{(\mathrm{a})}$ and $w_\mathrm{ms}^{(\mathrm{o})}$ denote the bandwidth needed by the MBS to serve MSUs when the RSBS is active and shut down, respectively.

Thus, the power consumption minimization problem can be formulated as follows.
\begin{subequations}\label{eq_Problem_RSBS}
    \begin{align}
    \mathcal{P}2:~~~& \min\limits_{I_\mathrm{R},\mu_\mathrm{E}} ~~ P_\mathrm{sum} \\
    \mbox{s.t.}~~~ & G_\mathrm{mm}\leq \eta, G_\mathrm{ss}\leq \eta, G_\mathrm{ms} \leq \eta, \\
                & 0 \leq w_\mathrm{mm} + w_\mathrm{ms}^{(\mathrm{o})} \leq W_\mathrm{m}, \\
                & 0\leq w_\mathrm{ss} \leq W_\mathrm{s},\\
    \end{align}
\end{subequations}
where the objective function is the total average on-grid power consumption, Eq.~(\ref{eq_Problem_RSBS}b) guarantees the QoS, Eq.~(\ref{eq_Problem_RSBS}c) and Eq.~(\ref{eq_Problem_RSBS}d) account for the bandwidth limitation of MBS and RSBS, respectively.

Similar to the single-HSBS case, there exists a tradeoff between the power consumption of the MBS and the handover cost.
By activating a RSBS for traffic offloading, the RF power of MBS is reduced, but handovers cause additional power consumption if the renewable energy is insufficient.
Besides, the energy consumption rate also has influences.
Specifically, increasing energy consumption rate enables the RSBS to serve more users, and thus reduce the RF power demand of the MBS.
However, higher energy consumption rate may also result in higher handover cost, as the energy queue may be emptied more frequently.
In what follows, we analyze this tradeoff relationship to solve problem $\mathcal{P}2$.
Denote by $\Delta_\mathrm{R}$ the power saving gain through activating the RSBS, given by
\begin{equation}\label{eq_gain_RSBS_definition}
    \Delta_\mathrm{R} = \{P_\mathrm{sum}|I_\mathrm{R}=0\} - \{P_\mathrm{sum}|I_\mathrm{R}=1\}.
\end{equation}
We summarize the relationship between $\Delta_\mathrm{R}$ and $\mu_\mathrm{E}$ in Proposition~1.

\emph{\textbf{Proposition~1.}} Denoting $\kappa = \zeta_\mathrm{EE} P_\mathrm{Cs} + \frac{\beta_\mathrm{m} P_\mathrm{Tm} R_\mathrm{Q}}{W_\mathrm{m} \tau_\mathrm{ms}}$, the power saving gain of a RSBS $\Delta_\mathrm{R}$ has following properties:
\begin{enumerate}
  \item $\Delta_\mathrm{R}$ increases linearly with $\mu_\mathrm{E}$ for $\mu_\mathrm{E}\leq \lambda_\mathrm{E}$;
  \item $\Delta_\mathrm{R}$ increases with $\mu_\mathrm{E}$ if $\kappa \geq 3 \lambda_\mathrm{E} C_\mathrm{HO}$ ;
  \item $\Delta_\mathrm{R}$ decreases with $\mu_\mathrm{E}$ for $\mu_\mathrm{E} > \lambda_\mathrm{E}$ if $\kappa \leq (1-\frac{1}{e}) \lambda_\mathrm{E} C_\mathrm{HO}$ ;
  \item If $(1-\frac{1}{e}) \lambda_\mathrm{E} C_\mathrm{HO} < \kappa < 3 \lambda_\mathrm{E} C_\mathrm{HO}$, $\Delta_\mathrm{R}$ is a concave function of $\frac{\lambda_\mathrm{E}}{\mu_\mathrm{E}}$ for $\mu_\mathrm{E} > \lambda_\mathrm{E}$, and the optimal condition is
      \begin{equation}\label{eq_opt_mu_RSBS}
        \lambda_\mathrm{E}C_\mathrm{HO} \frac{e^{-\frac{\lambda_\mathrm{E}}{\mu_\mathrm{E}}}}{\left(\frac{\lambda_\mathrm{E}}{\mu_\mathrm{E}}\right)^2} \left( -e^{-\frac{\lambda_\mathrm{E}}{\mu_\mathrm{E}}} +1+\frac{\lambda_\mathrm{E}}{\mu_\mathrm{E}}-\left(\frac{\lambda_\mathrm{E}}{\mu_\mathrm{E}}\right)^2 \right) =  \kappa.
      \end{equation}
\end{enumerate}
\emph{Proof}: Please refer to Appendix~\ref{appendix_RSBS_gain}.
\hfill \rule{4pt}{8pt}

Based on Eqs.~(\ref{eq_Problem_RSBS}b), (\ref{eq_Problem_RSBS}c) and $\varphi \leq 1$, $\mu_\mathrm{E}$ satisfies
\begin{equation}\label{eq_mu_condition}\footnotesize
    \frac{P_{\mathrm{Cs}}}{E} \leq \mu_\mathrm{E} \leq \frac{1}{E} \left\{ P_{\mathrm{Cs}}+ \min \left\{1, \frac{R_\mathrm{Q}}{\tau_\mathrm{ss} W_\mathrm{s}}\left(\rho_\mathrm{s}\pi D_\mathrm{s}^2 +1 \right) \right\} \beta_\mathrm{s} P_\mathrm{Ts} \right\},
\end{equation}
By combining Proposition~1 and Eq.~(\ref{eq_mu_condition}), the optimal energy consumption rate can be obtained.
The detailed result is omitted due to the space limitation.
In addition, the RSBS should be activated for traffic offloading if the maximal power saving gain is positive; otherwise, it should stay in OFF state.
Discussions are provided for some typical cases in Corollaries~3 and 4.

\textbf{Corollary~3.} If $\kappa \leq (1-\frac{1}{e}) \lambda_\mathrm{E} C_\mathrm{HO}$ (high handover cost) and $\lambda_\mathrm{E} \leq P_{\mathrm{Cs}}$, the RSBS should be deactivated.

\textbf{Corollary~4.} If $\lambda_\mathrm{E} E > P_{\mathrm{Cs}}+\beta_{s}P_\mathrm{Ts}$ (sufficient energy supply), the RSBS should be activated.
\section{Power Consumption Minimization for Multi-SBS Case}
    \label{sec_Multiple_SBS}
    In this section, we investigate the power consumption minimization problem for the case where multiple RSBSs, HSBSs, and CSBSs coexist.

\subsection{Problem Formulation}

When the multiple small cells coexist, the power consumption of the MBS is given by
\begin{equation}\label{eq_P_mbs_multi_sbs}
    P_\mathrm{MBS} = P_{\mathrm{Cm}} + \frac{\beta_\mathrm{m}P_\mathrm{Tm}}{W_\mathrm{m}} \left(w_\mathrm{mm} + \sum\limits_{n\in\mathfrak{B}} \tilde{w}_{\mathrm{ms},n}\right),
\end{equation}
where $\tilde{w}_{\mathrm{ms},n}$ is the average bandwidth needed by the MBSs to serve users in cell $n$.
Considering the characteristics of different SBSs, we have
\begin{equation}\label{eq_w_ms} \footnotesize
    \tilde{w}_{\mathrm{ms},n} \!=\! \left\{\! \begin{array}{l} \!w_{\mathrm{ms},n}^{(\mathrm{a})} I_{n} + w_{\mathrm{ms},n}^{(\mathrm{o})} (1- I_{n}),~\mbox{for HSBSs or CSBSs,} \\ \! I_n \left(\! w_{\mathrm{ms},n}^{(\mathrm{a})} (1\!-\!q_{0,n}) \!+\! w_{\mathrm{ms},n}^{(\mathrm{o})} q_{0n}\right) \!+\! w_{\mathrm{ms},n}^{(\mathrm{o})} (1\!-\! I_{n}),~\mbox{for RSBSs}, \end{array} \right.
\end{equation}
where $w_{\mathrm{ms},n}^{(\mathrm{a})}$ (and $w_{\mathrm{ms},n}^{(\mathrm{o})}$) is the bandwidth needed by the MSUs in cell $n$ when SBS$_n$ is active (and in off state), and $q_{0,n}$ is the probability that the battery of RSBS$_n$ is empty.

The total on-grid power consumption of the network is given by
\begin{equation}\label{eq_P_all_BS} \small
    P_{\mathrm{sum}} = P_\mathrm{MBS} \!+\! \sum\limits_{n\in \mathfrak{B}_\mathrm{R}} I_n P_{\mathrm{HO},n} \!+\! \sum\limits_{i\in \mathfrak{B}_\mathrm{H}} I_i P_{\mathrm{HSBS},i} \!+\! \sum\limits_{j\in \mathfrak{B}_\mathrm{C}} I_j P_{\mathrm{CSBS},j},
\end{equation}
where $P_{\mathrm{HO},n}$ reflects the handover cost of RSBS$_n$, while $P_{\mathrm{HSBS},i}$ and $P_{\mathrm{CSBS},j}$ denote the on-grid power consumption of the HSBS$_i$ and CSBS$_j$, respectively:
\begin{equation}\label{eq_P_HSBS_CSBS}
    \begin{split}
    P_{\mathrm{HSBS},i} & = q_{0,i} \left(P_{\mathrm{Cs},i} + \frac{w_{\mathrm{ss},i}}{W_\mathrm{s}} \beta_i P_{\mathrm{Ts},i}\right),\\
    P_{\mathrm{CSBS},j} & = P_{\mathrm{Cs},j} + \frac{w_{\mathrm{ss},j}}{W_\mathrm{s}} \beta_j P_{\mathrm{Ts},j}.
    \end{split}
\end{equation}

Then, the power minimization problem can be formulated as the following mixed integer programming problem:
\begin{subequations}\label{eq_problem_formulation}
    \begin{align}
       \mathcal{P}3:~~~&\min\limits_{\mathcal{I},\mu_\mathrm{E}}~~ P_\mathrm{sum} \\
       \mathrm{s.t.} ~~~& G_\mathrm{mm}\leq \eta , G_{\mathrm{ms},n}\leq \eta , G_{\mathrm{ss},n}\leq \eta ,~~n\in \mathfrak{B},\\
       & 0 \leq w_\mathrm{m}^{\max} \leq W_\mathrm{m},\\
       & 0\leq w_{\mathrm{ss},n}\leq W_\mathrm{s}, ~~ n\in \mathfrak{B}, \\
       & I_n \in \{0,1\},~~ n\in \mathfrak{B},
    \end{align}
\end{subequations}
where $w_\mathrm{m}^{\max}$ is the bandwidth needed at the MBS when all active RSBSs are shut down due to energy shortage:
\begin{equation}\label{eq_w_m_max}
    w_\mathrm{m}^{\max} = w_\mathrm{mm} + \sum\limits_{n\in (\mathfrak{B}_\mathrm{H}\bigcup \mathfrak{B}_\mathrm{C})} \tilde{w}_{\mathrm{ms},n} +  \sum\limits_{n\in\mathfrak{B}_\mathrm{R}} w_{\mathrm{ms},n}^{(\mathrm{o})}.
\end{equation}
Eq.~(\ref{eq_problem_formulation}b) guarantees the QoS of MMUs, MSUs and SSUs, while Eq.~(\ref{eq_problem_formulation}c) and (\ref{eq_problem_formulation}d) are due to the limited system bandwidth.

Based on the analytical results of the single-SBS cases, the optimal energy saving gain through offloading traffic to each individual SBS can be obtained, and the corresponding amount of traffic offloaded to the SBSs $\tilde{\mu}_\mathrm{E}$ is the optimal solution of problem $\mathcal{P}3$.
Then, the problem becomes which SBSs should be activated.
Now we consider the activation of SBSs.
Firstly, the SBSs with positive power saving gain should be activated, and the ON-OFF states is denoted as $\tilde{\mathcal{I}}$.
If $(\tilde{\mathcal{I}},\tilde{\mu}_\mathrm{E})$ is feasible for problem $\mathcal{P}3$, it is optimal since all SBSs whose activation reduces on-grid power consumption (i.e., the objective function) are activated.
Otherwise, the MBS is overloaded (i.e., Eq.~(\ref{eq_problem_formulation}c) does not hold) and more SBSs are selected to be reactivated, with the price of increasing the on-grid power consumption.
Therefore, the key problem is to determine which SBSs should be reactivated.

\subsection{SBS Reactivation and TEATO Scheme}

Note that the ON-OFF operation of RSBSs does not influence Eq.~(\ref{eq_problem_formulation}c).
The reason is that bandwidth $w_{\mathrm{ms},n}^{(\mathrm{o})}$ should be reserved no matter RSBS$_n$ is activated or not, since the MBS needs to serve the SSUs when RSBS$_n$ is shut down due to energy shortage.
Thus, activating new RSBSs only increases the total energy consumption without relieving the burden of MBS.
In contrast, offloading traffic to HSBSs and CSBSs can reduce the bandwidth requirement at MBS, according to Eq.~(\ref{eq_problem_formulation}c).
Therefore, only HSBSs and CSBSs should be considered for reactivation.

Denote $\mathfrak{B}_\mathrm{off}=\{1,2,...,N_\mathrm{off}\}$ the set of HSBSs and CSBSs in off-state according to $\tilde{\mathcal{I}}$, where $N_\mathrm{off}$ is the number of SBSs in OFF state.
Based on the analysis of Section~\ref{sec_Single_SBS}, once a HSBS or CSBS is activated, it should serve as many users as possible, and the optimal energy consumption rate is given by Eq.~(\ref{eq_mu_opt_HSBS}).
For each sleeping HSBS or CSBS $i$, activation increases the power consumption by $-\Delta_{i}$ (i.e., Eqs.~(\ref{eq_P_gain_HSBS_definition}) (\ref{eq_gain_RSBS_definition})), but reduces the bandwidth demand at MBS by $\delta_{i}= w_{\mathrm{ms},i}^{(\mathrm{o})}-w_{\mathrm{ms},i}^{(\mathrm{a})}$.
Thus, the SBS activating problem is a 0-1 knapsack problem as follows:
\begin{subequations}\label{eq_problem_formulation_2}
    \begin{align}
    \mathcal{P}4:~~~& \max\limits_{\mathcal{I}_\mathrm{a}} ~~\sum\limits_{i=1}\limits^{N_\mathrm{off}} I_{\mathrm{a},i} \Delta_{i}\\
    \mathrm{s.t.}~~~ & \sum\limits_{i=1}\limits^{N_\mathrm{off}} I_{\mathrm{a},i} \delta_{i} \geq \tilde{w}_\mathrm{m}^{\max}  - W_\mathrm{m},\\
                & I_{\mathrm{a},i} \in \{0,1\},~~i\in\mathfrak{B}_\mathrm{off},
    \end{align}
\end{subequations}
where $\tilde{w}_\mathrm{m}^{\max}$ is the bandwidth needed at the MBS under $(\tilde{\mathcal{I}},\tilde{\mu}_\mathrm{E})$.
The objective function is to minimize the increased power consumption caused by activating additional SBSs, Eq.~(\ref{eq_problem_formulation_2}b) guarantees that the MBS is not overloaded.

The 0-1 knapsack problem is NP-hard, and the optimal solution cannot be obtained within polynomial time.
Therefore, we relax $I_{\mathrm{a},i}$ to be a continuous variable $0 \leq \hat{I}_{\mathrm{a},i} \leq 1$, which denotes the probability to activate SBS $i$.
Then, problem $\mathcal{P}4$ becomes a linear programming function.
Applying the Lagrange multiplier method, we obtain the necessary condition of the optimal solution based on the Karush-Kuhn-Tucker conditions \cite{convex_optimization}, given as Proposition 2.

\emph{\textbf{Proposition~2}} The optimal solution of the relaxed problem $\hat{\mathcal{I}}_{\mathrm{a}}$ satisfies
\begin{enumerate}
  \item $\hat{I}_{\mathrm{a},i} = 1$ if $\frac{-\Delta_{i}}{\delta_{i}} < \nu$,
  \item $\hat{I}_{\mathrm{a},i} = 0$ if $\frac{-\Delta_{i}}{\delta_{i}} > \nu$,
  \item $\sum\limits_{i=1}\limits^{N_\mathrm{off}} \hat{I}_{\mathrm{a},i} \delta_{i} = \tilde{w}_\mathrm{m}^{\max}  - W_\mathrm{m},$
\end{enumerate}
where $\nu \leq 0$ is the Lagrange multiplier.
In fact, Proposition~2 indicates the SBSs which can offload traffic at lower extra power (i.e., smaller $\frac{-\Delta_{i}}{\delta_{i}}$) should be activated with higher priority.
Thereby, the suboptimal solution of problem $\mathcal{P}4$ is obtained, based on which $\mathcal{P}3$ can be solved.

With the Proposition~2, we propose the two-stage energy-aware traffic offloading (TEATO) scheme.
In the first stage, each SBS is analyzed independently to derive the maximal power saving gain $\Delta_i$ and the bandwidth relieved at the MBS $\delta_{i}$, same to that in the single-SBS case.
In the second stage, the ON-OFF states of SBSs are optimized, where the SBSs with positive power saving gain are activated first.
Then, additional HSBSs and CSBSs are reactivated if the MBS is still overloaded (Eq.~(\ref{eq_problem_formulation}c) does not hold), in the order of increasing $\frac{-\Delta_{i}}{\delta_{i}}$.
\section{Simulation Results}
    \label{sec_numerical_results}
    \begin{table}[!t]
        \caption{Simulation parameters}
        \label{tab_parameter}
        \centering
        \begin{tabular}{cc||cc}
        \hline
        \hline
        Parameter & Value & Parameter & Value \\
        \hline
        $D_\mathrm{m}$ & 1000m & $D_\mathrm{s}$ & 300m \\
        $\alpha_\mathrm{m}$ & 3.5 & $\alpha_\mathrm{s}$ & 4 \\
        $W_\mathrm{m}$ & 10MHz & $W_\mathrm{s}$ & 5MHz\\
        $R_\mathrm{Q}$ & 300kbps & $\eta$ & 0.05\\
        $\sigma^2$ & -105dBm/MHz & $\theta_\mathrm{m}$ & 1000\\
        $\theta_\mathrm{s}$ (single-SBS) & 500 & $\theta_\mathrm{s}$ (multi-SBS) & 2000\\
        \hline
        \hline
        \end{tabular}
\end{table}

\begin{figure}
  \centering
  \includegraphics[width=2.5in]{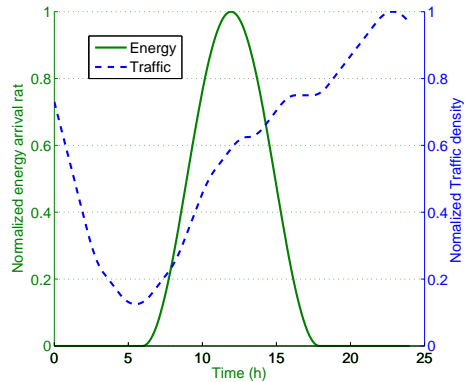}\\
  \caption{Daily traffic and energy profiles.}\label{fig_traffic_energy}
\end{figure}

In this section, we validate the accuracy of the derived outage probability, evaluate the energy saving gain of the optimal solution for the single-SBS case, and then demonstrate the performance of the proposed TEATO scheme for a HCN consisting of one MBS, one RSBS, one HSBS, and three CSBSs.
The SBSs are micro BSs, and the main simulation parameters can be found in Table~\ref{tab_parameter}.
Solar power harvesting devices are equipped at RSBS and HSBS.
Fig.~\ref{fig_traffic_energy} shows typical daily traffic and profiles.
The energy profile is based on real solar power generation data provided by the Elia group\footnote{The power generation is sampled and averaged every 15 minutes, and the data was collected in Belguim on August 1, 2014. For details, please refer to: http://www.elia.be/en/grid-data/power-generation/Solar-power-generation-data/Graph.}, and the adopted traffic profile proposed by the EARTH project has been widely used for performance evaluation \cite{JGong_TC} \cite{EARTH}.

    \begin{figure*}[!t]
        \centering
        \subfloat[MMU and SSU] {\includegraphics[width=2.5in]{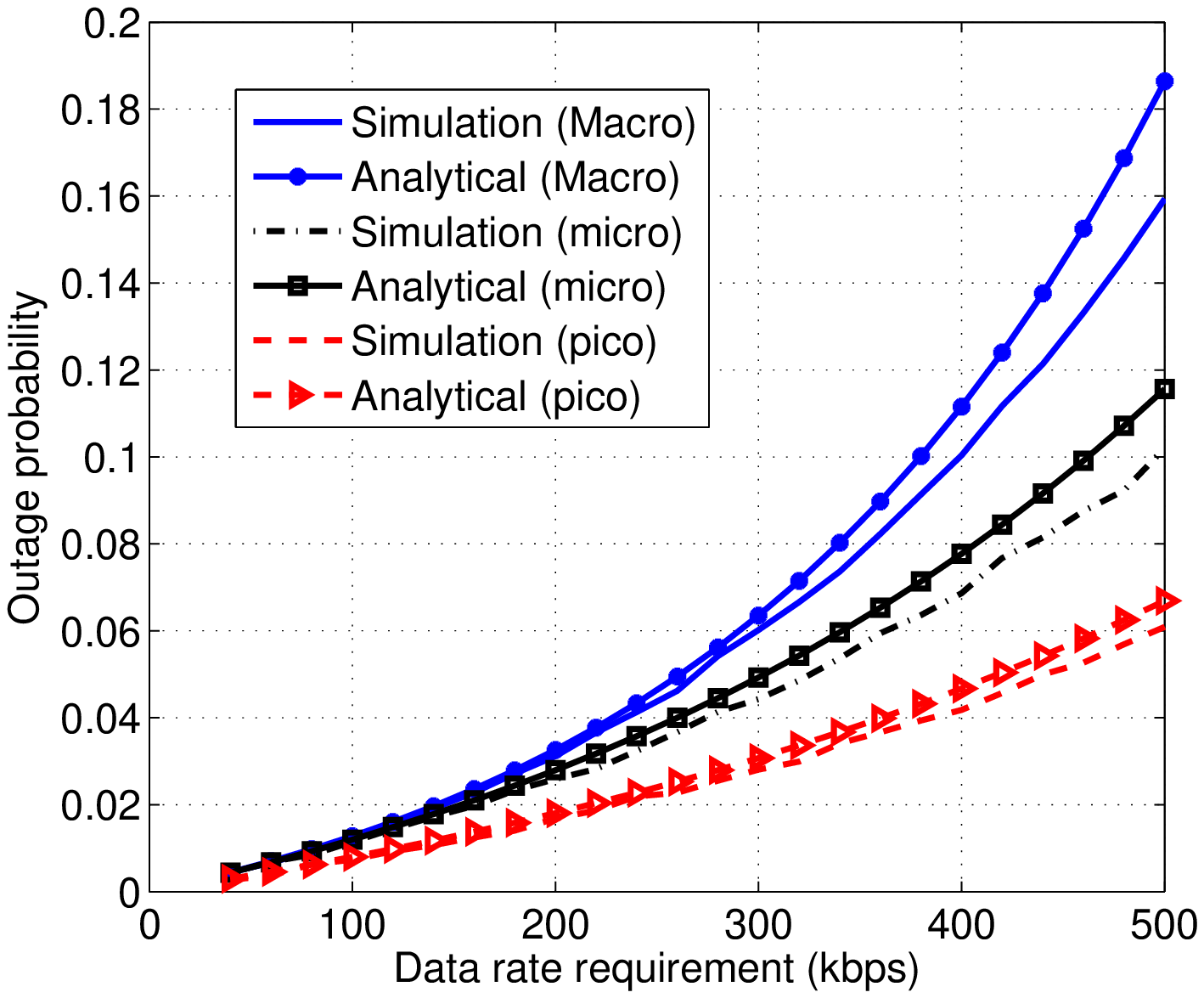}
        \label{fig_evaluation_MU_SU}}
        \hfil
        \subfloat[MSU]{\includegraphics[width=2.5in]{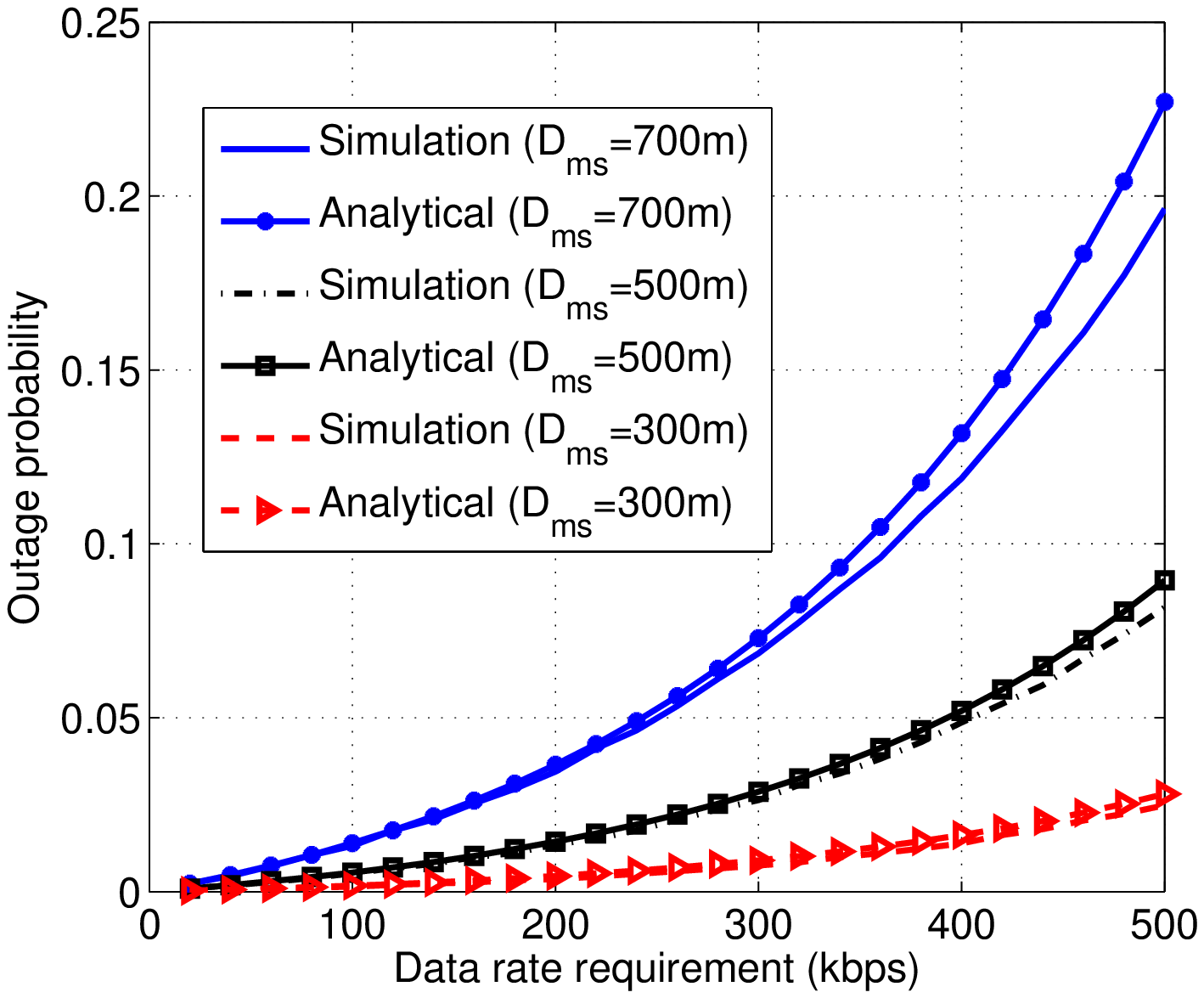}
        \label{fig_evaluation_MSU}}
        \caption{Outage probability.}
        \label{fig_evaluation}
    \end{figure*}

\subsection{Outage Probability Evaluation}

We evaluate the the analytical results in Theorems~1 and 2 via Monte Carlo simulations.
The number of users and their locations are randomly generated, and the results are averaged over 10000 simulation samples. Fig.~\ref{fig_evaluation}(a) compares the outage probability obtained by Theorem~1 with the simulation results, when the user density $\rho_\mathrm{m}$=20/km$^2$ and $\rho_\mathrm{s}$=70/km$^2$. We also consider the pico BS, whose coverage radius is 100m with user density 500/km$^2$.
Fig.~\ref{fig_evaluation}(b) compares the outage probability obtained by Theorem~2 with simulation results for different MBS-SBS distances, when the bandwidth of 3 MHz is shared by MSUs.
As shown in Fig.~\ref{fig_evaluation}, the analytical and simulation results generally match well, and both increase as the required data rate grows.
Although there exist certain approximation deviations, the analytical results are quite close to the simulation ones for small outage probabilities.
For example, the relative error rate is less than 10\% for $\eta<0.1$.
Therefore, Theorems~1 and 2 are applicable to the services with more strict QoS requirements, such as voice and real-time video streaming, whose typical outage probabilities are 0.02 or 0.05.

\subsection{Power Saving Gain of Single-SBS Case}

    \begin{figure*}[!t]
        \centering
        \subfloat[RSBS] {\includegraphics[width=2.5in]{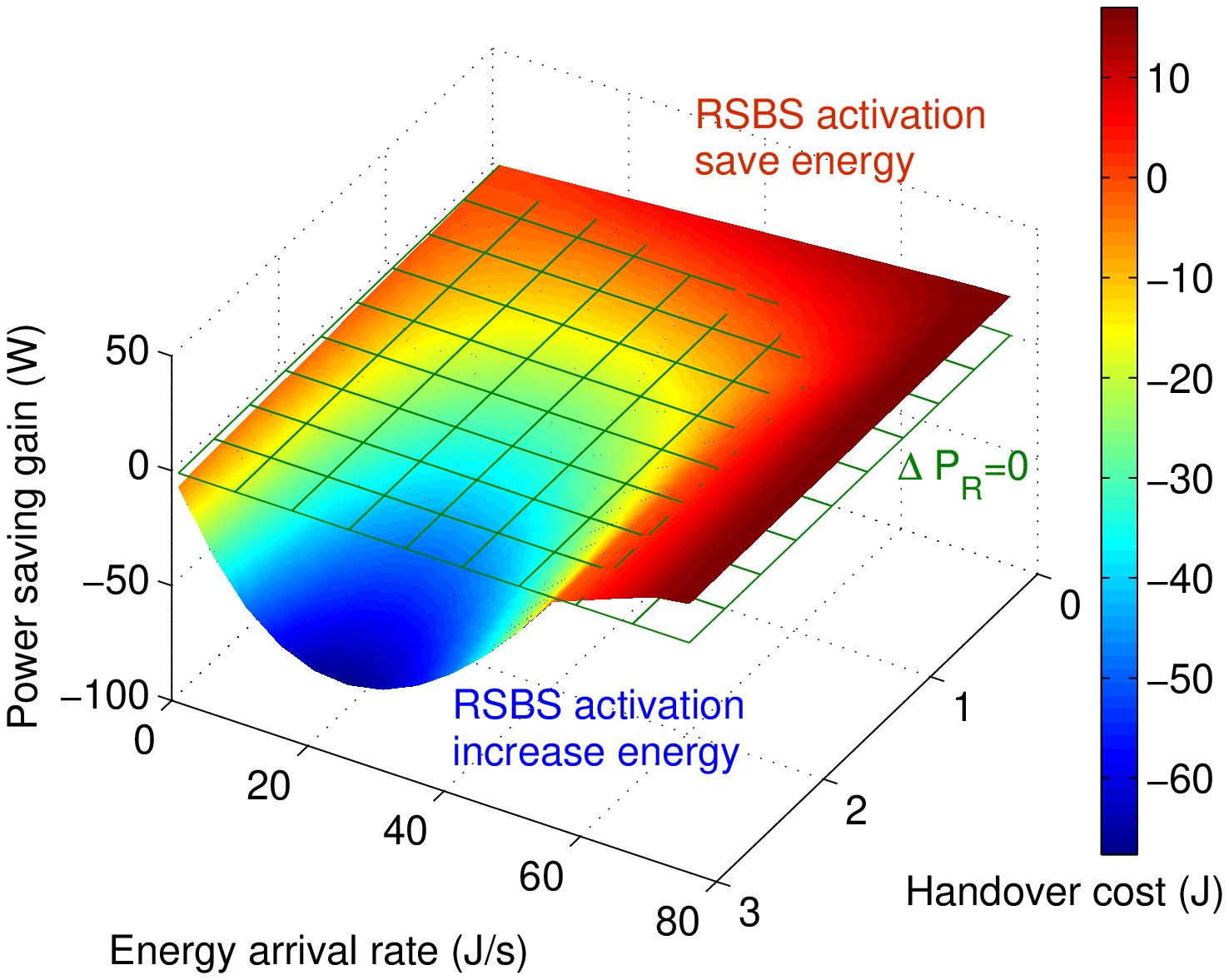}
        \label{fig_gain_RSBS}}
        \hfil
        \subfloat[HSBS]{\includegraphics[width=2.5in]{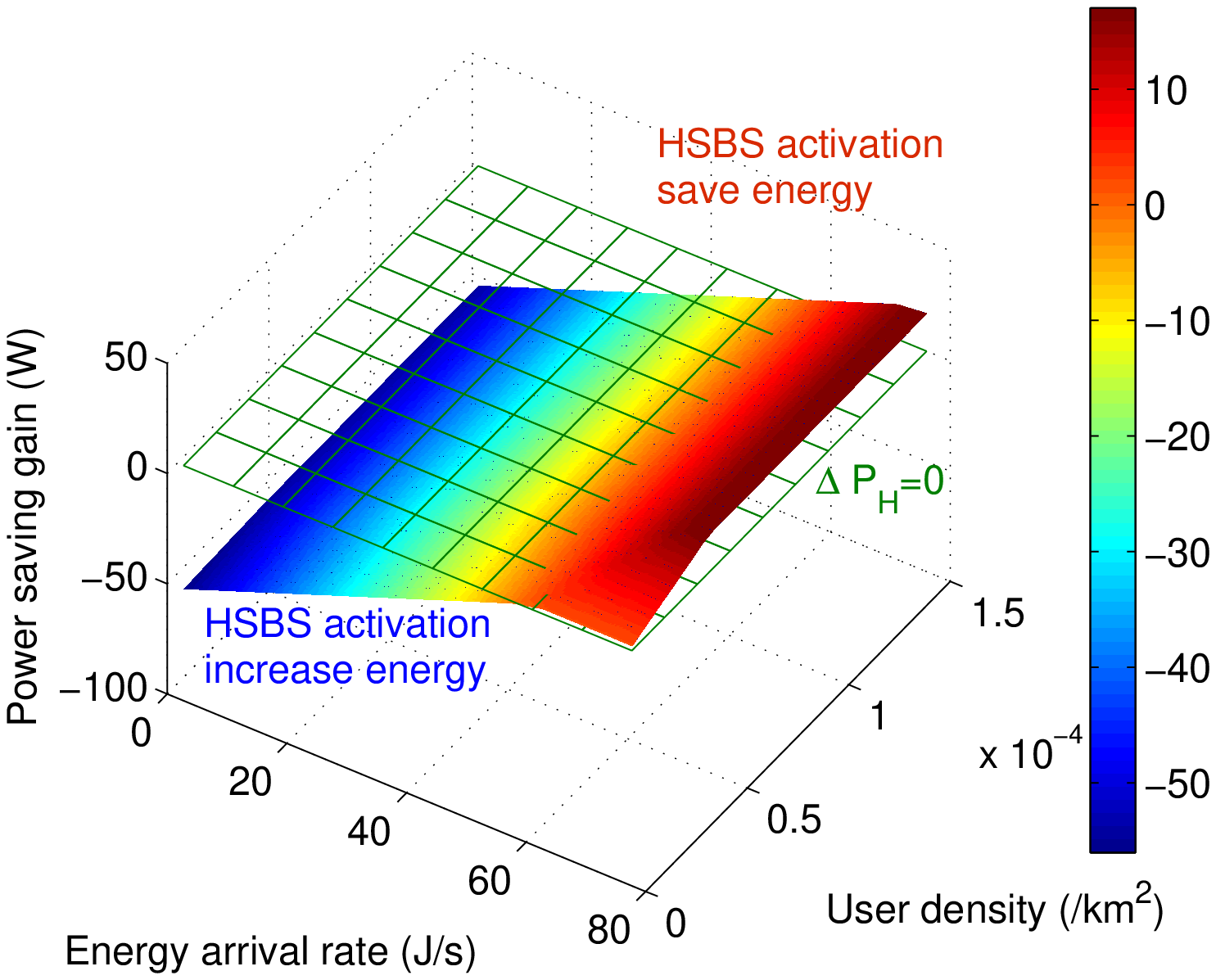}
        \label{fig_gain_HSBS}}
        \caption{Power saving gain of single SBS.}
        \label{fig_gain_single_SBS}
    \end{figure*}

Fig.~\ref{fig_gain_single_SBS} shows the maximal power saving gain achieved by activating the RSBS or the HSBS under different system parameters. Fig.~\ref{fig_gain_single_SBS}(a) shows the power saving gain for the RSBS.
It can be seen that power saving gain decreases with the increase of handover cost $C_\mathrm{ho}$, while it presents convexity with respect to the energy arrival rate.
Specifically, the power saving gain firstly decreases for $\lambda_\mathrm{E}<\lambda^\mathrm{th}_\mathrm{E}$, and then increases when $\lambda_\mathrm{E}>\lambda^\mathrm{th}_\mathrm{E}$.
Furthermore, $\lambda^\mathrm{th}_\mathrm{E}$ increases with the handover cost $C_\mathrm{HO}$.
Theoretically, the power saving gain is a convex function of the energy arrival rate when the renewable energy supply is insufficient, i.e., $\lambda_\mathrm{E}<\mu_\mathrm{E}$, which can be proved by taking the second derivative of Eq.~(\ref{eq_gain_RSBS}).
Intuitively, when the energy arrival rate $\lambda_\mathrm{E}$ is small, the transmission power saving gain of the MBS $\Delta_\mathrm{mbs}$ is very low due to insufficient green energy, and thus the total power saving gain is dominated by handover cost.
As the energy queue is empty most of the time for $\lambda_\mathrm{E}\rightarrow 0$, the handover frequency is approximately twice of the energy arrival rate, i.e., $P_\mathrm{HO}$ increases almost linearly as $\lambda_\mathrm{E}$ increases.
Therefore, the power saving gain decreases for $\lambda_\mathrm{E}<\lambda^\mathrm{th}_\mathrm{E}$.
As $\lambda_\mathrm{E}$ becomes large, the probability that the battery runs out decreases, and the $\Delta_\mathrm{mbs}$ balances $P_\mathrm{HO}$ out.
Accordingly, the total power saving gain is dominated by the saved transmission power, and thus increases with $\lambda_\mathrm{E}$.

Fig.~\ref{fig_gain_single_SBS}(b) shows the power saving gain for HSBSs, which increases with the density of users in the small cell. Moreover, it also increases with $\lambda_\mathrm{E}$ as the harvested energy reduces on-grid power consumption at the SBS. However, the power saving gain remains the same when the user density or $\lambda_\mathrm{E}$ is high, due to the limited service capability of the HSBS. In this case, the harvested energy is over-supplied and under-utilized. Note that the crossing line of two surfaces represents that the RF power saving gain balances out the constant power consumption of the SBS, which is easier to achieve with higher user density or energy arrival rate.

    \begin{figure*}[!t]
        \centering
        \subfloat[RSBS] {\includegraphics[width=2.5in]{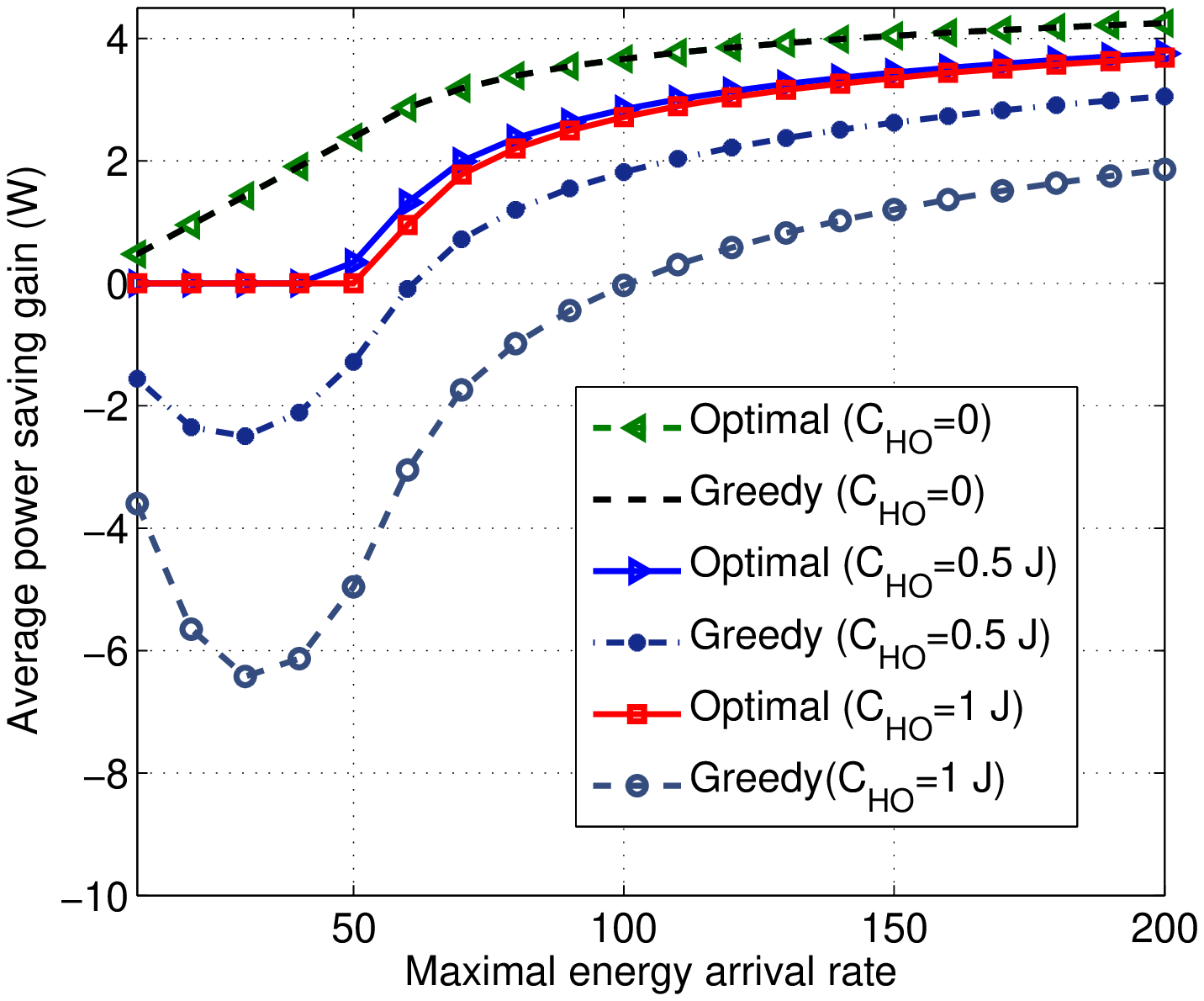}
        \label{fig_gain_RSBS_daily}}
        \hspace{2mm}
        \hfil
        \subfloat[HSBS]{\includegraphics[width=2.5in]{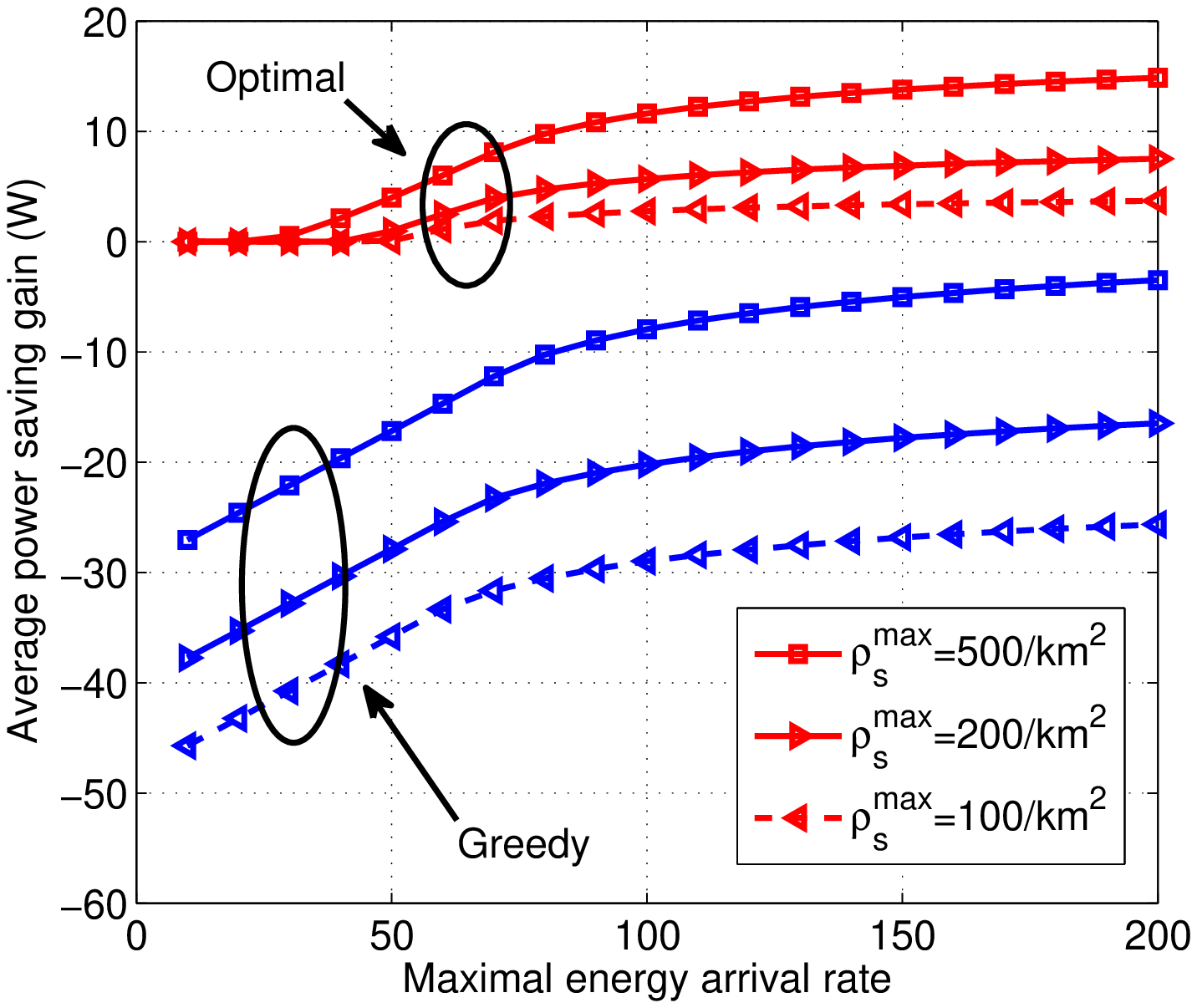}
        \label{fig_gain_HSBS_daily}}
        \caption{Average power saving gain of single SBS under daily traffic and energy profiles.}
        \label{fig_gain_single_SBS_daily}
    \end{figure*}

Fig.~\ref{fig_gain_single_SBS_daily} shows the power saving performance of the proposed optimal solution and the conventional greedy scheme for the daily traffic and energy arrival profiles from Fig.~\ref{fig_traffic_energy}. For the conventional greedy scheme, the RSBS or HSBS is always active to offload users as many as possible, whose intuition is to make use of the harvested energy as much as possible to avoid battery overflow. Fig.~\ref{fig_gain_single_SBS_daily}(a) shows the power saving gain with respect to the maximum energy arrival rate for the RSBS, when the maximal user density of the day is $\rho_\mathrm{s} = 100/$km$^2$.
Under the greedy scheme, the power saving gain firstly decreases and then increases with the energy arrival rate due to the handover cost, same as the results in Fig.~\ref{fig_gain_single_SBS}(a). Whereas the power saving gain of the optimal solution increases monotonously with the energy arrival rate, as the amount of offloading traffic and the ON-OFF states of SBS are jointly optimized to avoid the frequent handover.
In addition, the two schemes achieve the same performance if the handover cost can be ignored ($C_\mathrm{HO}=0$). However, the performance of the greedy scheme degrades significantly as the handover cost increases, and the average power saving gain even becomes negative when the energy arrival rate is low. In other words, deploying a RSBS may increase the total power consumption with the greedy scheme.

Fig.~\ref{fig_gain_single_SBS_daily}(b) shows the comparison of the power saving performance of the proposed optimal solution and the greedy scheme for HSBSs under different traffic loads. It can be seen that the average power saving gain degrades as the traffic load decreases. Moreover, the power saving gain of the greedy scheme even becomes negative for low energy arrival rate (i.e., cloudy days).
Whereas, the proposed optimal solution always guarantees positive power saving gain.

\subsection{Network Power Saving Gain}

\begin{figure}[t]
	\centering
	\includegraphics[width=2.5in]{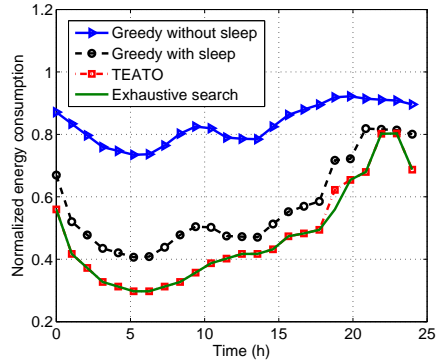}
	\caption{Daily power consumption.}\label{fig_network_gain_daily}
\end{figure}

For the multi-SBS case, the maximal user density in macro cell is set as 20 /km$^2$, while the maximal energy arrival rate is set as 200 J/s.
Suppose that the user density in small cells is twice as that in the macro cell. Fig.~\ref{fig_network_gain_daily} shows the normalized network power consumption for different schemes when the handover cost per time is 2 J.
The results are normalized by the power consumption of the HCN consisting of 1 MBS and 4 CSBSs, where no cell sleeping and power control is adopted. 
The difference between the two greedy schemes is whether CSBSs can go into sleep during low traffic hours for energy saving.
It can be seen that the greedy scheme without cell sleeping achieves higher power saving gain when the energy arrival rate is high.
In addition, the greedy scheme with cell sleeping can further improve the performance, especially when the energy arrival rate is high and the traffic load is low.
This is because the constant power consumption is reduced by turning off CSBSs.
Moreover, the TEATO scheme can achieve the best performance, as it adjusts the RF power and avoids activating SBSs which may bring negative power saving gain.
The performance with optimal solution of problem $\mathcal{P}3$ is demonstrated as the green solid line, which is obtained by exhaustive search.
Notice that the performance of the proposed scheme TEATO is the same as the optimal solution for most of the time, validating the 0-1 relaxation of $\mathcal{P}3$.

\begin{figure}[t]
	\centering
	\includegraphics[width=2.5in]{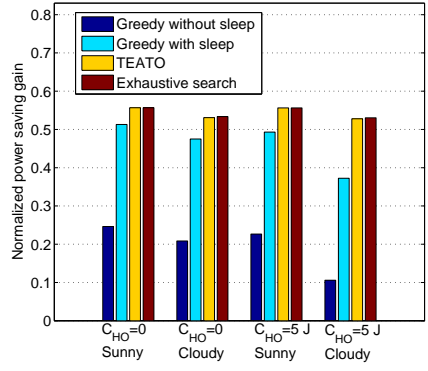}
	\caption{Average power saving gain.}\label{fig_network_gain_daily_bar}
\end{figure}

Fig.~\ref{fig_network_gain_daily_bar} shows the average power saving gain of the three schemes under different weather conditions, where the maximal energy arrival rates for sunny and cloudy days are set as 500 W and 50 W, respectively. It can be seen that the proposed TEATO can save around 50\% energy for all scenarios compared with the greedy scheme without cell sleeping.
Besides, the performance of the greedy schemes degrades when the handover cost increases in cloudy days.
The results of Figs.~\ref{fig_network_gain_daily} and \ref{fig_network_gain_daily_bar} demonstrate the benefits and necessity to conduct energy-aware power control and dynamic cell sleeping.

\section{Conclusions and Future Work}
    \label{sec_conclusions}

    In this paper, we have investigated the on-grid power saving gain through offloading traffic for green
    heterogeneous networks.
    The analytical results reflect the conversion rate of harvested energy into on-grid power through traffic offloading, and also offer insights for practical green cellular  network operations, e.g., whether the SBS should be activated, and how much traffic should be offloaded to the SBS such that the on-grid power can be minimized.
    Furthermore, an energy-efficient traffic offloading scheme, namely TEATO, has been proposed for the multi-SBSs case.
    Simulation results have been given to demonstrate that TEATO can reduce about 50\% of power consumption on average for daily traffic and renewable energy profiles, compared with the greedy schemes.
    For the future work, the traffic offloading among overlapped SBSs will be studied, and differentiated services with diverse QoS requirements will be considered.


\appendices{}
   \section{Proof of Theorem~1}
    \label{appendix_SBS}

        Based on Eq.~(\ref{eq_SINR_SBS}), we have
        \begin{subequations} \footnotesize
            \label{eq_appendix_MBS_outage_Nm}
            \begin{align}
                & \mathds{P} \left\{\gamma_{\mathrm{ss},n} \geq 2^{(K_{\mathrm{ss},n}+1)\frac{R_\mathrm{Q}}{ w_{\mathrm{ss},n}}} - 1 \right\} \nonumber \\
                &= \int_0^{D_{n}} \mathds{P} \left\{ h_{n} \geq \frac{(\theta_\mathrm{s}+1)\sigma^2 {W_\mathrm{s}} }{P_{\mathrm{Ts},n} d^{-\alpha_\mathrm{s}} } \left(2^{(K_{\mathrm{ss},n}+1) \frac{R_\mathrm{Q}}{w_{\mathrm{ss},n}}} -1 \right)\right\} \frac{2d}{D_{n}^2} \mathrm{d} d  \nonumber \\
                &= \int_0^{D_{n}} \exp\left( -\frac{(\theta_\mathrm{s}+1)\sigma^2 {W_\mathrm{s}} }{P_{\mathrm{Ts},n} d^{-\alpha_\mathrm{s}} } \left(2^{(K_{\mathrm{ss},n}+1) \frac{R_\mathrm{Q}}{w_{\mathrm{ss},n}}} -1 \right)\right) \frac{2d}{D_{n}^2} \mathrm{d} d
                \label{eq_appendix_MBS_rayleigh} \\
                &= \int_0^{D_{n}} \left( 1-\frac{(\theta_\mathrm{s}+1)\sigma^2 {W_\mathrm{s}} }{P_{\mathrm{Ts},n} d^{-\alpha_\mathrm{s}} } \left(2^{(K_{\mathrm{ss},n}+1) \frac{R_\mathrm{Q}}{w_{\mathrm{ss},n}}} -1 \right)\right) \frac{2d}{D_{n}^2} \mathrm{d} d  \label{eq_appendix_MBS_PDF_rate} \\
                &= 1 - \frac{2D_{n}^{\alpha_\mathrm{s}} W_\mathrm{s} }{\alpha_\mathrm{s}+2} \frac{(\theta_\mathrm{s}+1)\sigma^2}{P_{\mathrm{Ts},n}} \left( 2^{(K_{\mathrm{ss},n}+1) \frac{R_\mathrm{Q}}{w_{\mathrm{ss},n}}}-1 \right), \nonumber
            \end{align}
        \end{subequations}
    where Eq.~(\ref{eq_appendix_MBS_rayleigh}) is based on the definition of Rayleigh fading, and (\ref{eq_appendix_MBS_PDF_rate}) holds as $\frac{P_{\mathrm{Ts},n}}{(\theta_\mathrm{s}+1)\sigma^2 W_\mathrm{s}} \rightarrow \infty$, based on $\lim\limits_{x \rightarrow 0} e^{-x}=1-x$.

    Recall that the probability distribution of $K_{\mathrm{ss},n}$ follows Poisson distribution.
    By substituting Eq.~(\ref{eq_appendix_MBS_outage_Nm}) into Eq.~(\ref{eq_outage_SBS}), the outage probability of a typical SBS user is:
        \begin{equation} \footnotesize
            \begin{split}
                & G_{\mathrm{ss},n} \!=\! 1\! -\! \sum_{K=0}^\infty \mathds{P} \left( \gamma_{\mathrm{ss},n} \geq 2^{(K+1) \frac{R_\mathrm{Q}}{w_{\mathrm{ss},n}}} -1 \right) P_{K_{\mathrm{ss},n}}(K) \\
                &\!=\! 1\!-\!\sum_{K=0}^\infty \mathds{P} \left( \gamma_{\mathrm{ss},n} \geq 2^{(K+1) \frac{R_\mathrm{Q}}{w_{\mathrm{ss},n}}} -1 \right) \frac{(\pi D_{n}^2 \rho_{n})^K}{K!}e^{- \pi D_{n}^2 \rho_{n}} \\
                &= \frac{2 D_{n}^{\alpha_\mathrm{s}} (\theta_\mathrm{s}\!+\!1)\sigma^2 W_\mathrm{s} }{P_{\mathrm{Ts},n} \left(\alpha_\mathrm{s}\!+\!2\right) } \! \left( 2^{\frac{R_\mathrm{Q}}{w_{\mathrm{ss},n}}}\exp\left(\pi D_{n}^2 \rho_{n} \left( 2^{\frac{R_\mathrm{Q}}{w_{\mathrm{ss},n}}} \!-\! 1 \right)\right)\!-\!1\right).
            \end{split}
        \end{equation}
    As $\lim\limits_{x\rightarrow0} \frac{a^x-1}{x} = \ln a$ for $a>1$, we have $\lim\limits_{\frac{R_\mathrm{Q}}{w_{\mathrm{ss}}} \rightarrow 0} 2^{\frac{R_\mathrm{Q}}{w_{\mathrm{ss},n}}}-1 = \ln2 \cdot \frac{R_\mathrm{Q}}{w_{\mathrm{ss},n}}$.
    Hence, Theorem~1 is proved.

\section{Proof of Theorems~3 and 4}
\label{appendix_HSBS}

With Eqs.~(\ref{eq_P_total_HSBS}-\ref{eq_sum_power_HSBS_detail_2}), the power saving gain by activating the HSBS $\Delta_\mathrm{H}$ is given by
\begin{equation}\label{eq_gain_hsbs}
    \Delta_\mathrm{H} = \frac{w_\mathrm{ms}^{(\mathrm{o})}-w_\mathrm{ms}^{(\mathrm{a})}}{W_\mathrm{m}} \beta_\mathrm{m} P_\mathrm{Tm} - q_0 (P_{\mathrm{Cs}} + \frac{w_\mathrm{ss}}{W_\mathrm{s}} \beta_s P_\mathrm{Ts}).
\end{equation}
To satisfy the outage probability constraints from Eq.~(\ref{eq_Problem_HSBS}b), we have
\begin{subequations}\label{eq_RSBS_msa_mso}
    \begin{align}
    w_\mathrm{ms}^{(\mathrm{o})} & \geq \frac{R_\mathrm{Q}}{\tau_\mathrm{ms}} \left( 1+ \rho_\mathrm{s} \pi {D_\mathrm{s}}^2 \right), \\
    w_\mathrm{ms}^{(\mathrm{a})} & \geq \frac{R_\mathrm{Q}}{\tau_\mathrm{ms}} \left( 1+ (1-\varphi) \rho_\mathrm{s} \pi {D_\mathrm{s}}^2 \right),
    \end{align}
\end{subequations}
based on Theorem~2.
$\varphi$ is offloading ratio constrained by the service capability of SBS (Eq.~(\ref{eq_outage_SBS_simple})):
\begin{equation}\label{eq_ratio_offloaded}
    \varphi \leq \frac{\frac{\tau_{\mathrm{ss}} w_\mathrm{ss} }{R_\mathrm{Q}}-1}{\rho_\mathrm{s} \pi {D_\mathrm{s}}^2},
\end{equation}
where $w_\mathrm{ss}$ depends on the energy consumption rate of the HSBS $\mu_\mathrm{E}$ given by
\begin{equation}\label{eq_RSBS_ws}
    w_\mathrm{ss} = \frac{1}{\beta_\mathrm{s} W_\mathrm{s}} \left(\mu_\mathrm{E} E - P_{\mathrm{Cs}}\right).
\end{equation}
Note that $q_0=0$ when $\mu_\mathrm{E}\leq \lambda_\mathrm{E}$; otherwise, $q_0 = 1-\frac{\lambda_{E}}{\mu_\mathrm{E}}$.
Substitute Eq.~(\ref{eq_RSBS_msa_mso})-(\ref{eq_RSBS_ws}) into Eq.~(\ref{eq_gain_hsbs}), Theorem~3 is proved.

According to Eq.~(\ref{eq_gain_hsbs_2}), the power saving gain increases with $\mu_\mathrm{E}$ if $ \zeta_\mathrm{EE} \geq 1 $; otherwise, it achieves the maximum when $\mu_\mathrm{E}=\lambda_\mathrm{E}$.
As $\zeta_\mathrm{EE}>1$ in real systems, the optimal solution $\tilde{\mu}_\mathrm{E}$ takes its maximal feasible value.
Based on Eqs.~(\ref{eq_Problem_HSBS}e), and $\varphi \leq 1$,  we have
\begin{equation}\label{eq_w_s_HSBS_range}
    0\leq w_\mathrm{ss} \leq \min \left\{W_\mathrm{s}, \frac{R_\mathrm{Q}}{\tau_\mathrm{ss}}\left(\rho_\mathrm{s}\pi D_\mathrm{s}^2 +1 \right) \right\},
\end{equation}
Recall the relationship between $\mu_\mathrm{E}$ and $w_\mathrm{ss}$ in Eq.~(\ref{eq_Problem_HSBS}c), and Theorem~4 is proved.

\section{Proof of Proposition~1}
\label{appendix_RSBS_gain}

$\Delta_\mathrm{R}$ can be written as
\begin{equation}\label{eq_gain_RSBS}
    \Delta_\mathrm{R} = (1 - q_0) \frac{w_\mathrm{ms}^{(\mathrm{o})}-w_\mathrm{ms}^{(\mathrm{a})}} {W_\mathrm{m}} \beta_\mathrm{m} P_\mathrm{Tm} - P_\mathrm{HO}
    \triangleq \Delta_\mathrm{mbs} - P_\mathrm{HO},
\end{equation}
where $\Delta_\mathrm{mbs}$ represents the power saved at the MBS.
Based on Eqs.~(\ref{eq_RSBS_msa_mso}-\ref{eq_RSBS_ws}), we have
\begin{equation}\label{eq_gain_MBS_g} \footnotesize
    \Delta_\mathrm{mbs} = \left\{ \begin{array}{ll} \zeta_\mathrm{EE} \mu_\mathrm{E} E - \left( \zeta_\mathrm{EE} P_\mathrm{Cs} + \frac{\beta_\mathrm{m} P_\mathrm{Tm} R_\mathrm{Q}}{W_\mathrm{m} \tau_\mathrm{ms}} \right), &  \mu_\mathrm{E}\leq \lambda_\mathrm{E} \\
    \zeta_\mathrm{EE} \lambda_\mathrm{E} E - \frac{\lambda_\mathrm{E}}{\mu_\mathrm{E}} \left( \zeta_\mathrm{EE} P_\mathrm{Cs} + \frac{\beta_\mathrm{m} P_\mathrm{Tm} R_\mathrm{Q}}{W_\mathrm{m} \tau_\mathrm{ms}} \right), & \mu_\mathrm{E} > \lambda_\mathrm{E}
    \end{array} \right. ,
\end{equation}
which reflects the conversion rate of harvested energy into on-grid power.
Substituting $\Delta_\mathrm{mbs}$ and $P_\mathrm{HO}$ in Eq.~(\ref{eq_gain_RSBS}) with Eqs.~(\ref{eq_gain_MBS_g}) and (\ref{eq_RSBS_switching}), the power saving gain $\Delta_\mathrm{R}$ can be obtained.

When $\mu_\mathrm{E}\leq\lambda_\mathrm{E}$, $P_\mathrm{HO}=0$, and $\Delta_\mathrm{R} = \Delta_\mathrm{mbs}$ increases linearly with $\mu_\mathrm{E}$.
When $\mu_\mathrm{E}>\lambda_\mathrm{E}$, denote $x = \frac{\lambda_\mathrm{E}}{\mu_\mathrm{E}}$ where $x\in(0,1)$.
The first derivation of $\Delta_\mathrm{R}$ with respect to $x$ is
\begin{equation}\label{appendix_diff_delta_g}\small
    \begin{split}
    & \frac{\mathrm{d} \Delta_\mathrm{R}}{\mathrm{d} x} = - \kappa - \left[ (1-\frac{1}{x^2})(1-e^{-x}) +(\frac{1}{x}-1)e^{-x} \right] \lambda_\mathrm{E} C_\mathrm{HO}\\
    & = - \kappa - \lambda_\mathrm{E} C_\mathrm{HO} \frac{e^{-x}}{x^2} \left( -e^\mathrm{x} +1+x-x^2 \right) \triangleq - \kappa + \lambda_\mathrm{E} C_\mathrm{HO} f(x).
    \end{split}
\end{equation}
Now we analyze the property of $f(x)$. As
\begin{subequations}\label{appendix_fx}
    \begin{align}
    \frac{\mathrm{d}f(x)}{\mathrm{d}x} & = \frac{2 e^{-x}}{x^3} \left( -e^\mathrm{x} +1+x+\frac{x^2}{2} - \frac{x^3}{2}  \right)\\
    & = \frac{2 e^{-x}}{x^3} \left( - \frac{x^3}{2} - \sum\limits_{i=3}\limits^{\infty} \frac{x^i}{i!}  \right) <0,
    \end{align}
\end{subequations}
and $0<x<1$, we have $ 1-e^{-1}<f(x)<\frac{3}{2}$.
In addition, $\Delta_\mathrm{R}$ is a concave function of $x$ as $\frac{\mathrm{d}^2 \Delta_\mathrm{R}}{\mathrm{d} x^2} = f'(x)<0$. Therefore, $\frac{\mathrm{d} \Delta_\mathrm{R}}{\mathrm{d} x} < 0$ for $\kappa \geq 3\lambda_\mathrm{E}C_\mathrm{HO}$, and $\frac{\mathrm{d} \Delta_\mathrm{R}}{\mathrm{d} x} > 0$ for $\kappa \leq (1-\frac{1}{e}) \lambda_\mathrm{E}C_\mathrm{HO}$.
Otherwise, there exists $\tilde{x}$ satisfying $ \frac{\mathrm{d} \Delta_\mathrm{R}}{\mathrm{d} x}|_{\tilde{x}} = 0$, and the corresponding energy consumption rate $\mu_\mathrm{E} = \frac{\lambda_\mathrm{E}}{\tilde{x}}$ maximizes the power saving gain. Hence, Proposition~1 is proved.

\end{document}